\documentclass[11pt]{article}

\usepackage{graphicx,ifthen,color,algorithm,palatino}
\usepackage{amsmath,amsthm,amssymb,amsfonts,verbatim}
\usepackage{mathrsfs,accents,setspace}
\usepackage{subfigure,caption}
\usepackage{multirow,makecell}
\usepackage{placeins}
\usepackage[authoryear]{natbib}
\usepackage{hyperref}
\usepackage{soul}
\newboolean{ShowFigures}
\newboolean{UnBlinded}
\newboolean{DoubleSpaced}
\setboolean{ShowFigures}{true}
\setboolean{UnBlinded}{true}
\setboolean{DoubleSpaced}{true}
\def\ba{\boldsymbol{a}}

\def\br{\boldsymbol{r}}

\def\bv{\boldsymbol{v}}
\def\bw{\boldsymbol{w}}
\def\bx{\boldsymbol{x}}
\def\by{\boldsymbol{y}}

\def\bA{\boldsymbol{A}}
\def\bB{\boldsymbol{B}}

\def\bI{\boldsymbol{I}}

\def\bQ{\boldsymbol{Q}}

\def\balpha{\boldsymbol{\alpha}}
\def\bbeta{\boldsymbol{\beta}}

\def\bdeta{\boldsymbol{\eta}}
\def\btheta{\boldsymbol{\theta}}

\def\blambda{\boldsymbol{\lambda}}
\def\bmu{\boldsymbol{\mu}}
\def\bnu{\boldsymbol{\nu}}

\def\bpsi{\boldsymbol{\psi}}

\def\bSigma{\boldsymbol{\Sigma}}

\def\vecof{\mbox{vec}}
\def\tr{\mbox{tr}}

\def\simind{\stackrel{{\tiny \mbox{ind.}}}{\sim}}

\begin{document}

\ifthenelse{\boolean{DoubleSpaced}}{\setstretch{1.5}}{}

\centerline{\Large\bf Variational Bayes for Mixture of Gaussian}
\vskip4mm
\centerline{\Large\bf Structural Equation Models}
\vskip7mm
\ifthenelse{\boolean{UnBlinded}}{
\centerline{\normalsize\sc By Khue-Dung Dang$\null^{1}$\footnotemark, Luca Maestrini$\null^{2}$\footnotemark[\value{footnote}] and Francis K.C. Hui$\null^{2}$}
\footnotetext{These authors equally contributed to this work.}
\vskip5mm
\centerline{\textit{$\null^1$University of Melbourne and $\null^2$Australian National University}}}{}
\vskip6mm
\centerline{\today}

\vskip6mm

\centerline{\large\bf Abstract}
\vskip2mm

{
Structural equation models (SEMs) are commonly used to study the structural relationship between observed variables and latent constructs. Recently, Bayesian fitting procedures for SEMs have received more attention thanks to their potential to facilitate the adoption of more flexible model structures, and variational approximations have been shown to provide fast and accurate inference for Bayesian analysis of SEMs. However, the application of variational approximations is currently limited to very simple, elemental SEMs. We develop mean-field variational Bayes algorithms for two SEM formulations for data that present non-Gaussian features such as skewness and multimodality. The proposed models exploit the use of mixtures of Gaussians, include covariates for the analysis of latent traits and consider missing data. We also examine two variational information criteria for model selection that are straightforward to compute in our variational inference framework. The performance of the MFVB algorithms and information criteria is investigated in a simulated data study and a real data application. 

}

\vskip3mm
\noindent
\textit{Keywords:} approximate inference; latent variables; mean field variational Bayes; missing data; variational information criteria.

\section{Introduction}

Multivariate data such as those arising from behavioural, educational, medical and social sciences studies can be analysed through structural equation models (SEMs).
These models are commonly used to test hypotheses regarding the structural relationship between predictive factors of interest and multiple observable outcomes by means of latent factors \citep{kaplan2008structural}. For example, this article considers a study on child growth and development where investigators are interested in assessing the impact of environmental factors such as prenatal alcohol exposure on a child cognitive function. 
``Cognition'', or ``cognitive function'' is a complex and multi-faceted construct that is difficult to measure directly and social scientists typically rely on a battery of different but related tests to provide a measure of general mental ability.  
The resulting cognition scores can be represented by latent factors which can in turn be modelled as a function of various predictors of interest. 

In our motivation study, investigators are interested in assessing the impact of prenatal alcohol exposure on cognitive function of children. \cite{jacobson2004maternal} and \cite{dang2023bsem} used simple SEM formulations based on a Gaussian response assumption and a simple latent factor structure to study the dose-response relationship between prenatal alcohol exposure and child cognition. In particular, \cite{dang2023bsem} adopted a Bayesian SEM to analyse the effect of prenatal alcohol exposure on cognition. However, their model was not able to capture some prominent features of the outcomes such as skewness and multimodality. Here we make use of mixtures of Gaussians to better capture the structure of the data and improve model fitting. The application of our framework is not limited to our motivating example but is relevant to the general class of data that are typically examined using SEMs, which often exhibit non-Gaussian structures. We focus on a model where each outcome is assumed to follow a mixture of Gaussians and also show that this formulation is preferrable to other proposals assuming a mixture of Gaussians on the latent variables.

Bayesian analysis of SEMs is receiving increasing attention as frequentist approaches such as those based on maximum likelihood estimation and weighted least squares may encounter computational and theoretical problems for small sample sizes or non-Gaussian data as in the case of the data examined in this work. Bayesian formulations of SEMs facilitate incorporation of useful information via prior distributions and can accommodate more flexible model structures, such as those with crossed-loadings \citep{lee2007structural}. Another important advantage of Bayesian approaches is the ability to get information on non-Gaussian features such as multimodality in marginal densities \citep{arhonditsis2006exploring}.
Markov chain Monte Carlo (MCMC) methods are typically used for fitting and inference of Bayesian SEMs but they may suffer from slow convergence and long running times, making them less appealing when it comes to performing tasks such as tuning of priors, sensitivity analysis or model selection. \cite{dang2022fitting} investigated the use of variational approximations for SEMs and showed that this approximation technique can provide fast and reliable fitting for basic SEMs. Recent works have also envisages a growing interest in the application of variational inference to structural equation modelling \citep{zhang2022model,kaplan2023bayesian,fazio2024gaussian}.

In this work, we build upon the base variational approximation framework outlined by \cite{dang2022fitting} and demonstrate the use of variational Bayes to solve a wider range of problems modelled through SEMs. We consider models that contemplate the presence of missing data, include covariates to explain latent factors and, more importantly, that make use of mixtures of Gaussian densities on the outcomes. This is different from more common model formulations used to deal with non-standard features of outcomes or latent variables such as mixtures of SEMs \citep[e.g.,][]{vermunt2005structural}, infinite mixtures for nonparametric modelling of latent traits \citep[e.g.,][]{dunson2006bayesian} or skewed distributions \citep[e.g.,][]{asparouhov2016structural}. We also consider an alternative model where a Gaussian mixture is applied to the latent variables and the conditional distribution of the outcomes is assumed to be Gaussian. We then utilize a variational version of the Akaike information criterion and propose a variational formulation of the Watanabe-Akaike information criterion to perform model selection. Through the application of these criteria we demonstrate that the model including a Gaussian mixture on the responses better captures the features of the real data in exam, besides requiring simpler prior hyperparameter tuning and initialisation of the variational algorithms compared to the model where the mixture is placed on the latent factors. Our simulation experiments also demonstrate that the variational information criteria are effective tools for selecting a reliable model for inference on the main parameters of interests in SEM analysis. 

In Section \ref{sec:baseModel} we formulate the main model of interest with a Gaussian mixture on the responses and an alternative formulation where the mixture is placed on the latent factors rather than the outcomes. In Section \ref{sec:modelComp} we introduce tools for performing model comparison in the context of variational inference. Section \ref{sec:numStudy} provides a numerical study to assess the models through the proposed selection criteria. The analysis of the prenatal alcohol exposure data is provided in Section \ref{sec:application}. Conclusions and future directions are outlined in Section \ref{sec:conclusion}.


%
\section{Structural equation models with Gaussian-mixture outcomes\label{sec:baseModel}}


Consider a set of latent factors $\eta_i$ associated to individuals indexed by $i=1,\ldots,N$ and each regressed over a $p\times 1$ vector of covariates $\bx_i$ and corresponding vector of coefficients $\bbeta$ of equal length. Let $y_{ij}$ denote the $j$th outcome, with $j=1,\ldots,M$, observed for the $i$th individual and assumed to be conditionally distributed according to a Gaussian mixture of $H_j$ components. The proposed model is
\begin{equation}
    \begin{array}{c}
       y_{ij}\,\vert\,\mu_{jh},\lambda_j,\eta_i,\psi^2_{jh}\simind \sum_{h=1}^{H_j} w_{jh} N\big(\mu_{jh}+\lambda_j\eta_i,\psi^2_{jh}\big), \\[1ex]
       \eta_i\,\vert\,\bbeta,\sigma^2\simind N(\bx_i^T\bbeta,\sigma^2),\\[1ex] 
        \mu_{jh}\simind N(\mu_{\mu_{jh}},\sigma^2_{\mu_{jh}}),\,\,\,\lambda_j\simind N(\mu_{\lambda},\sigma^2_\lambda),\,\,\,\psi_{jh}^2\simind\mbox{Inverse-Gamma}(\alpha_{\psi^2},\beta_{\psi^2}),\,\,\,\\[1ex] 
        \bbeta\sim N(\bmu_{\bbeta},\bSigma_{\bbeta}),\quad\sigma^2\sim\mbox{Inverse-Gamma}(\alpha_{\sigma^2},\beta_{\sigma^2}),\quad 
        \bw_{j}\sim\mbox{Dirichlet}(H_j,\alpha_w),
    \end{array}
\label{eq:mainSEM}
\end{equation}
for $i=1,\ldots,N$, $j=1,\ldots,M$ and $h=1,\ldots,H_j$. Here $\mbox{Dirichlet}(H_j,\alpha_w)$ denotes a symmetric Dirichlet distribution of order $H_j$ with concentration parameter $\alpha_w>0$. Note that we allow for the hyperparameters $\mu_{\mu_{jh}}$ and $\sigma^2_{\mu_{jh}}$ of the Gaussian priors on $\mu_{jh}$ to be different and this will be motivated later on in our application. Our study involves a single unobserved component, cognitive function, therefore the latent factors $\eta_i$ are univariate. Nonetheless, the model above could be easily extended to accommodate for multiple unobserved components (see \citeauthor{dang2022fitting}, \citeyear{dang2022fitting}, Section 7). The vectors $\bx_i$ include individual measurements of covariates (in our study, alcohol exposure and a propensity score) that affect the latent variables and are assumed to not contain an intercept term. Consistently with the features of the data in exam, the number of mixture components $H_j$ can vary across the $M$ outcomes so that different outcomes can potentially be modelled with different numbers of mixture components or with a Gaussian distribution when $H_j=1$.
To facilitate fitting, model \eqref{eq:mainSEM} can be rewritten in terms of auxiliary variables $\ba_{ij}=(a_{ij1},\ldots,a_{ijH_j})^T$ as
\begin{equation}
\begin{array}{c}
    p(y_{ij}\,\vert\,\mu_{jh},\lambda_j,\eta_i,\psi^2_{jh},\ba_{ij}) = \prod_{h=1}^{H_j} \left[\psi_{jh}^{-1}(2\pi)^{-1/2} \exp \left\{-  \dfrac{(y_{ij}-\mu_{jh}-\lambda_j\eta_i)^2}{2\psi_{jh}^2}\right\}\right] ^{a_{ijh}}, \\[1.5ex]
    \ba_{ij} \simind \mbox{Multinomial}(1; \bw_j),\,\,\, \eta_i\,\vert\,\bbeta,\sigma^2\simind N(\bx_i^T\bbeta,\sigma^2),\\[1ex]
    \mu_{jh}\simind N(\mu_{\mu},\sigma^2_{\mu}),\,\,\,\lambda_j\simind N(\mu_{\lambda},\sigma^2_\lambda),\,\,\,\psi_{jh}^2\simind\mbox{Inverse-Gamma}(\alpha_{\psi^2},\beta_{\psi^2}),\\[1ex] 
    \bbeta\sim N(\bmu_{\bbeta},\bSigma_{\bbeta}),\quad\sigma^2\sim\mbox{Inverse-Gamma}(\alpha_{\sigma^2},\beta_{\sigma^2}),\quad 
    \bw_{j}\sim\mbox{Dirichlet}(H_j,\alpha_w).
    \end{array}
\label{eq:mainSEM2}
\end{equation}

In many real data applications including the one considered in this work, some outcomes may be unobserved for certain individuals. Assuming that the data are missing at random, one straightforward way to account for this type of missing observations is to use a full information maximum likelihood approach, which corresponds to taking only the contribution to the likelihood from the observed responses into account \citep{arbuckle1996full,finkbeiner1979estimation,merkle2021efficient}. This can be implemented by introducing the set $\mathcal{S}_{\tiny\mbox{obs}}$ of $(i,j)$ pairs corresponding to outcomes $y_{ij}$ that are observed and rewriting model \eqref{eq:mainSEM} in terms of the outcomes $y_{ij}$ such that $(i,j)\in\mathcal{S}_{\tiny\mbox{obs}}$. In a similar way, the likelihood function can be written by replacing products over all pairs of $i=1,\ldots,N$ and $j=1,\ldots,M$ with products over all $(i,j)\in\mathcal{S}_{\tiny\mbox{obs}}$. 
Given that the elements of $\by_i$ are independent conditionally on the latent variables, the likelihood function arising from the conditional distribution of the outcomes in \eqref{eq:mainSEM2} under the missing data situation described above can be written as $$p(\by|\bmu,\blambda,\bdeta,\bpsi^2,\ba) = \prod_i \prod_{j: (i,j)\in\mathcal{S}_{\tiny\mbox{obs}} } p(y_{ij}|\mu_{ij},\lambda_j,\eta_i,\psi_{jh}^2,\ba_{ij}), $$
which reflects the fact we have at least one outcome observed for each individual $i$.

\begin{figure}
    \centering
    \includegraphics[width=0.7\linewidth]{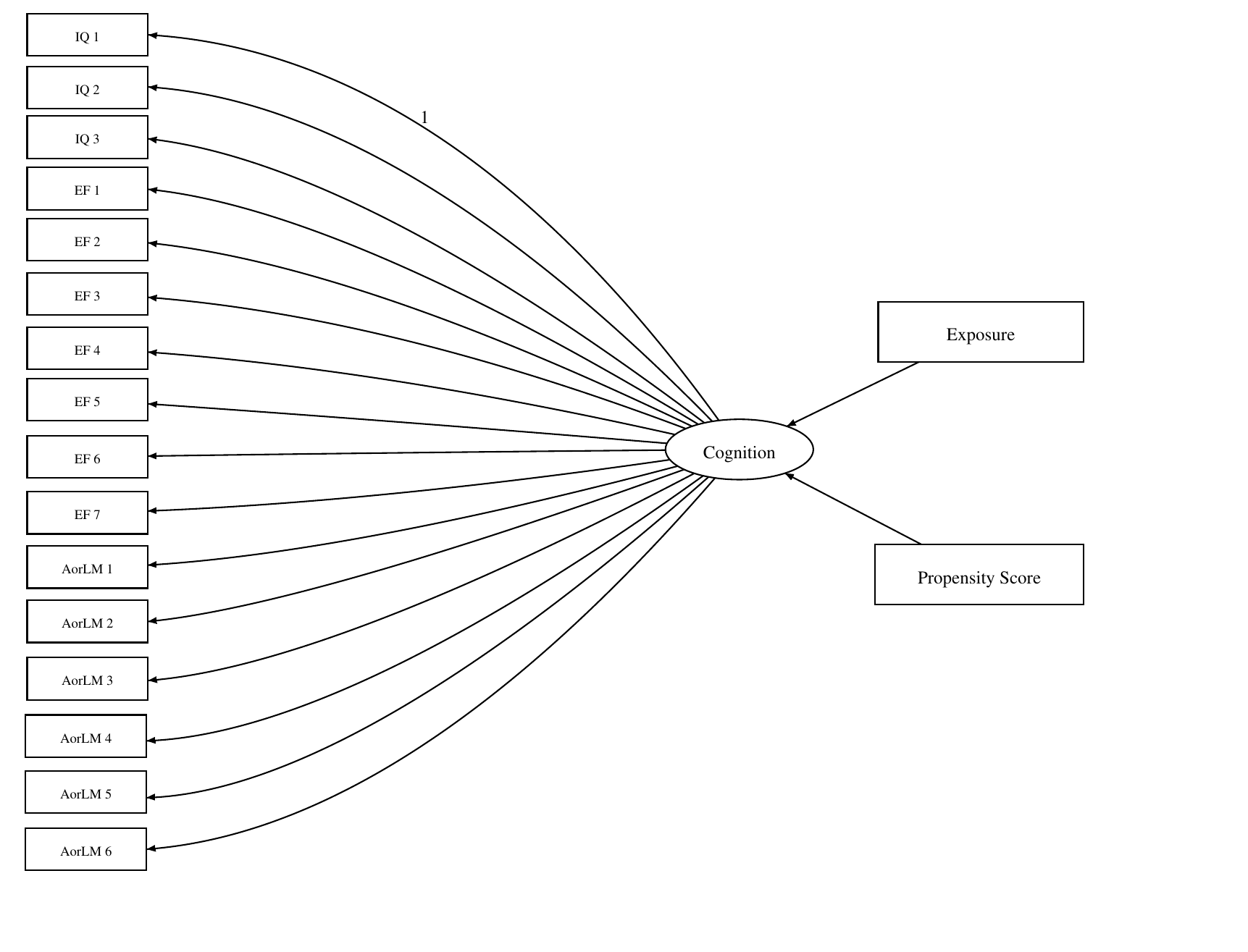}
    \caption{The path diagram for both models \eqref{eq:mainSEM} and \eqref{eq:altSEM} in the context of the cognitive data study including 16 outcomes and 2 covariates (`Exposure' and `Propensity Score'). The outcomes `IQ 1' to `AorLM 6' are described in Section \ref{sec:application}. The number 1 over the arrow pointing to `IQ 1' means that the factor loading connecting the latent variable `Cognition' and the outcome `IQ 1' is set to 1.}
    \label{fig:pathDetroit}
\end{figure}

In the literature of SEMs and their applications, the relationship between observed outcomes and latent variables is often represented through a path diagram. Figure \ref{fig:pathDetroit} shows the path diagram for our cognitive data examined through model \eqref{eq:mainSEM}. The observed outcomes are represented as rectangles to which the arrows departing from the only latent variable, cognition (represented as an ellipsis) are pointing. The diagram also shows the two exogenous covariates that are supposed to affect cognition: a propensity score and alcohol exposure. A more detailed description of the data is provided in Section \ref{sec:application}.  

\subsection{Model fitting via VB}

\cite{dang2022fitting} outlined a framework for fast approximate fitting of SEMs via mean field variational Bayes (MFVB) which we also employ in this work to fit model \eqref{eq:mainSEM} (specifically its auxiliary variable version \eqref{eq:mainSEM2}) and the alternative model proposed later. The motivating idea is that a faster model fitting technique facilitates tasks such as prior elicitation or selection of the number of Gaussian mixture components. As for mainstream variational approximations, MFVB boils down to finding an approximating density $q$ for which the Kullback–Leibler divergence between the approximating density itself and the posterior density function is minimized, subject to convenient restrictions on $q$.
If $\by$ is a generic vector of data, $\btheta\in\Theta$ represents all model parameters and $q$ is an arbitrary density function defined over $\Theta$, the logarithm of the marginal likelihood satisfies
\begin{equation}
\log p(\by)\geq\log\underline{p}(\by;q)\equiv\int q(\btheta)\log\left\{\frac{p(\by,\btheta)}{q(\btheta)}\right\}d\btheta,
\label{eq:LowerBound}
\end{equation}
where $\underline{p}(\by;q)$ is a marginal likelihood lower-bond depending on $q$. It can easily be shown that maximizing this lower-bound is equivalent to minimizing the Kullback-Leibler divergence
\begin{equation*}
\mbox{KL}(q(\btheta)\,\Vert\, p(\btheta\vert\by))=\int q(\btheta)\log\left\{\frac{q(\btheta)}{p(\btheta\vert\by)}\right\}d\btheta.
\end{equation*}
If the approximating density is factorized according to a partition $(\btheta_1,\ldots,\btheta_K)$ of $\btheta$ such that $q(\btheta)=\prod_{k=1}^Kq(\btheta_k)$, then the optimal approximating densities satisfy
\begin{equation}
	q^*(\btheta_k)\propto\exp\big[E_{q(\btheta\backslash \btheta_k)}\{\log p(\btheta_k\vert\by,\btheta\backslash \btheta_k )\}\big],\quad k=1,\ldots,K,
\label{eq:qkupdate}
\end{equation}
where $E_{q(\btheta\backslash \btheta_k)}$ denotes the expectation with respect to all the approximating densities except $q(\btheta_k)$ and $\btheta\backslash \btheta_k$  represents the entries of $\btheta$ with $\btheta_k$ omitted.
It can also be shown that the maximization of the lower-bound can be performed via a coordinate ascent scheme converging to a local maximizer of the lower bound under mild regularity conditions. A more detailed introduction to MFVB as presented above is provided, for example, in Section 2.2 of \cite{ormerod2010explaining}.

As an illustration, consider the full conditional density of $\mu_{jh}$, for $j=1,\ldots,M$ and $h=1,\ldots,H_j$, arising from the log likelihood function of model \eqref{eq:mainSEM}:
\begin{equation*}
    p(\mu_{jh}\,\vert\,\mbox{rest})\propto\exp\left[-\frac{\mu_{jh}^2}{2}\left(\sum_{i:(i,j)\in\mathcal{S}_{\tiny\mbox{obs}}}\frac{a_{ijh}}{\psi_{jh}^2}+\frac{1}{\sigma^2_{\mu_{jh}}}\right)+\mu_{jh}\left\{\sum_{i:(i,j)\in\mathcal{S}_{\tiny\mbox{obs}}}\frac{a_{ijh}(y_{ij}-\lambda_j\eta_i)}{\psi_{jh}^2}+\frac{\mu_{\mu_{jh}}}{\sigma^2_{\mu_{jh}}}\right\}\right].
\end{equation*}
Furthermore, consider the following factorization of the variational approximating density:
\begin{align}
\begin{split}
    &q(\bmu,\blambda,\bdeta,\bpsi^2,\bbeta,\sigma^2,\ba,\bw)=q(\bmu)q(\blambda)q(\bpsi^2)q(\bbeta)q(\sigma^2)\prod_{i=1}^Nq(\eta_i)q(\ba)q(\bw)\\
    &\qquad\qquad=q(\sigma^2)q(\bbeta)\prod_{j=1}^M\prod_{h=1}^{H_j} \{q(\psi_{jh}^2)q(\mu_{jh})\}\prod_{j=2}^M q(\lambda_j)\prod_{i=1}^Nq(\eta_i)\prod_{(i,j)\in\mathcal{S}_{\tiny\mbox{obs}}}q(\ba_{ij})\prod_{j=1}^M q(\bw_j).
\end{split}
    \label{eq:simpSEMcovariatesRestr}
\end{align}
From application of \eqref{eq:qkupdate} and given the factorization in \eqref{eq:simpSEMcovariatesRestr}, the optimal approximating density for $\mu_{jh}$
\begin{equation}
\begin{array}{c}
    q^*(\mu_{jh})\quad\mbox{is}\quad N\big(\mu_{q(\mu_{jh})},\sigma^2_{q(\mu_{jh})}\big),\quad j=1,\ldots,M,\,\,h=1,\ldots,H_j,\\[3ex]
    \mbox{with}\quad\mu_{q(\mu_{jh})}=\sigma^2_{q(\mu_{jh})}\left\{\sum_{i:(i,j)\in\mathcal{S}_{\tiny\mbox{obs}}}\mu_{q(a_{ijh})}\mu_{q(1/\psi_{jh}^2)}\big(y_{ij}-\mu_{q(\lambda_j)}\mu_{q(\eta_i)}\big)+\dfrac{\mu_{\mu_{jh}}}{\sigma^2_{\mu_{jh}}}\right\}\\[3ex]
    \mbox{and}\quad\sigma^2_{q(\mu_{jh})}=\left(\sum_{i:(i,j)\in\mathcal{S}_{\tiny\mbox{obs}}}\mu_{q(a_{ijh})}\mu_{q(1/\psi_{jh}^2)} + \dfrac{1}{\sigma^2_{\mu_{jh}}} \right)^{-1},
\end{array}
\label{eq:q_mujh}
\end{equation}
where $\mu_{q(a_{ijh})}$, $\mu_{q(1/\psi_{jh}^2)}$, $\mu_{q(\lambda_j)}$ and $\mu_{q(\eta_i)}$ are defined in Algorithm \ref{alg:MFVBmixtureSEMwithLambdaetaAndCovariatesNew}.
The optimal approximating densities of the remaining parameters can be derived in a similar manner and have the following forms:
\begin{equation*}
\begin{array}{c}
    q^*(\bbeta)\quad\mbox{is}\quad N\left(\bmu_{q(\bbeta)},\bSigma_{q(\bbeta)}\right),\\[1ex] 
    q^*(\lambda_j)\quad\mbox{is}\quad N\big(\mu_{q(\lambda_j)},\sigma^2_{q(\lambda_j)}\big),\quad j=2,\ldots,M,\\[1ex]
    q^*(\eta_i)\quad\mbox{is}\quad N\big(\mu_{q(\eta_i)},\sigma^2_{q(\eta_i)}\big),\quad i=1,\ldots,N,\\[1ex]
    q^*(\psi_{jh}^2)\quad\mbox{is}\quad \mbox{Inverse-Gamma}\Big(\alpha_{q(\psi_{jh}^2)},\beta_{q(\psi_{jh}^2)}\Big),\quad j=1,\ldots,M,\,\,h=1,\ldots,H_j,\\[1ex]
    q^*(\sigma^2)\quad\mbox{is}\quad\mbox{Inverse-Gamma}\left(\alpha_{q(\sigma^2)},\beta_{q(\sigma^2)}\right),\\[1ex]
    q^*(\ba_{ij})\quad\mbox{is}\quad\mbox{Multinomial}\big(1;\mu_{q(\ba_{ij})}\big),\quad i=1,\ldots,N,\quad k=1,\ldots, K,\\[1ex]
    q^*(\bw_j)\quad\mbox{is}\quad \mbox{Dirichlet}(H_j,\balpha_{q(\bw_j)}),\quad j=1,\ldots,M,
\end{array}
\end{equation*}
where $G\left([\begin{array}{cc}
    \bv_1^T\,\,
    \bv_2^T
    \end{array}]^T;\bQ,\br,s\right)\equiv-(1/8)\tr\left(\bQ\{\vecof^{-1}(\bv_2)\}^{-1}[\bv_1\bv_1^T\{\vecof^{-1}(\bv_2)\}^{-1}-2\bI]\right)$\\ $-(1/2)\br^T\{\vecof^{-1}(\bv_2)\}^{-1}\bv_1-(1/2)s$, for a $d\times 1$ vector $\bv_1$, $d^2\times 1$ vector $\bv_2$ such that $\vecof^{-1}(\bv_2)$ is symmetric, $d\times d$ matrix $\bQ$, $d\times 1$ vector $\br$ and $s\in\mathbb{R}$, as defined in \cite{wand2017fast}, and the quantities on which the approximating densities parameters depend are given in Algorithm \ref{alg:MFVBmixtureSEMwithLambdaetaAndCovariatesNew}. Following standard practice in the SEM literature, we fix $\lambda_1$ to $1$ and introduce the constraint $\lambda_j>0$ to ensure model identifiability. 

An important aspect of the proposed modelling framework is the choice of the number of Gaussian mixture components $H_j$. This can be performed, for instance, resorting to software such as the package `mixtools' \citep{mixtools} in R \citep{R2024} to cluster the measurements of each outcome and then using the estimated number of clusters as $H_j$. Visual inspection of the empirical density plots of outcome could also help identifying or confirming the number of mixture components.  

%
\begin{algorithm}[htb]
	\begin{center}
		\begin{minipage}[t]{154mm}
			\begin{small}
				\begin{itemize}
					\setlength\itemsep{2pt}
					\item[] \textbf{Data Input:} $\by_i$, $i=1,\ldots,N$, vectors of length $M$.
					\item[] \textbf{Hyperparameter Input:} $\mu_{\mu_{jh}},\mu_{\lambda}\in\mathbb{R}$, $\sigma_{\mu_{jh}}, \sigma_\lambda,\alpha_{\psi^2},\beta_{\psi^2},\alpha_{\sigma^2},\beta_{\sigma^2}\in\mathbb{R}^+$, $\alpha_w=1$, $\bmu_{\bbeta}\in\mathbb{R}^p$, $\bSigma_{\bbeta}$ being a $p\times p$ symmetric positive definite matrix.
					\item[] \textbf{Initialize:}  $\mu_{q(\mu_{jh})},\mu_{q(\log(\psi^2_{jh}))} \in \mathbb{R}$, $\mu_{q(\mu_{jh}^2)},\mu_{q(1/\psi_{jh}^2)}\in\mathbb{R}^+$, $j=1,\ldots,M$, $h=1,\ldots,H_j$; $\mu_{q(\lambda_j)}\in \mathbb{R},\mu_{q(\lambda_j^2)}\in \mathbb{R}^+$, $j=2,\ldots,M$; $\mu_{q(\eta_i)}\in \mathbb{R},\mu_{q(\eta_i^2)}\in \mathbb{R}^+$, $i=1,\ldots,N$; $\mu_{q(1/\sigma^2)}\in\mathbb{R}^+$, $\bmu_{q(\bbeta)}\in\mathbb{R}$, $\mu_{q(\log(w_{jh}))}\in \mathbb{R}^{-}$, $j=1,\ldots,M$, $h=1,\ldots,H_j$; $\alpha_{q(\sigma^2)}=\frac{N+2\alpha_{\sigma^2}}{2}$, $\mu_{q(\lambda_1)}=\mu_{q(\lambda_1^2)}=1$.
					\item[] \textbf{Cycle until convergence:}
					\begin{itemize}
						\setlength\itemsep{2pt}

						\item[] For $j = 1,\ldots,M$:
						\begin{itemize}
						\setlength\itemsep{2pt}
						\item[] For $i=1,\ldots,N$, $h = 1,\ldots,H_j$:
						\begin{itemize}
						    \setlength\itemsep{2pt}
                        \item[] $\tau_{ijh} \longleftarrow \mu_{q(\log(w_{jh}))}-\frac{1}{2}\mu_{q(\log(\psi_{jh}^2))}-\frac{1}{2}\log(2\pi) -\frac{1}{2}\mu_{q(1/\psi_{jh}^2)}\big(y_{ij}^2+\mu_{q(\mu^2_{jh})}$ \item[]\qquad\qquad$+\mu_{q(\lambda_j^2)}\mu_{q(\eta_i^2)}-2y_{ij}\mu_{q(\mu_{jh})}-2y_{ij}\mu_{q(\lambda_j)}\mu_{q(\eta_i)}+2\mu_{q(\mu_{jh})}\mu_{q(\lambda_j)}\mu_{q(\eta_i)}\big)$
    			        \end{itemize}
						\item[] For $i=1,\ldots,N$, $h = 1,\ldots,H_j$: $\mu_{q(a_{ijh})} \longleftarrow \exp(\tau_{ijh}) \Big / \sum_{h=1}^{H_j} \exp(\tau_{ijh})$
    			        \item[] If $j>1$:
    			        \begin{enumerate}
    			            \item[] $\sigma^2_{q(\lambda_j)}\longleftarrow\left( \sum_{i=1}^N\sum_{h=1}^{H_j} \mu_{q(a_{ijh})}\mu_{q(\eta_i^2)}\mu_{q(1/\psi_{jh}^2)}+\frac{1}{\sigma^2_\lambda} \right)^{-1}$
    			        \item[] $\mu_{q(\lambda_j)}\longleftarrow\sigma^2_{q(\lambda_j)}\left\{\sum_{i=1}^N\sum_{h=1}^{H_j}\mu_{q(a_{ijh})}\mu_{q(\eta_i)}\mu_{q(1/\psi_{jh}^2)}\big(y_{ij} -\mu_{q(\mu_{jh})}\big)+\frac{\mu_{\lambda}}{\sigma^2_{\lambda}}\right\}$
    			        \item[] $\mu_{q(\lambda_j^2)}\longleftarrow\sigma^2_{q(\lambda_j)}+\mu_{q(\lambda_j)}^2$
    			        \end{enumerate}
                   	  \item[] For $h = 1,\ldots,H_j$: 
                            \begin{enumerate}
                        \item[] $\sigma^2_{q(\mu_{jh})}\longleftarrow\left(\sum_{i=1}^N\mu_{q(a_{ijh})}\mu_{q(1/\psi_{jh}^2)} + 1/\sigma^2_{\mu_{jh}} \right)^{-1}$

                        \item[] $\mu_{q(\mu_{jh})}\longleftarrow\sigma^2_{q(\mu_{jh})}\left\{\sum_{i=1}^N\mu_{q(a_{ijh})}\mu_{q(1/\psi_{jh}^2)}\big(y_{ij}-\mu_{q(\lambda_j)}\mu_{q(\eta_i)}\big)+\mu_{\mu_{jh}}/\sigma^2_{\mu_{jh}}\right\}$    
                        \item[] $\mu_{q(\mu_{jh}^2)} \longleftarrow \sigma^2_{q(\mu_{jh})} + \mu^2_{q(\mu_{jh})}$
    			        \item[] $\alpha_{q(\psi_{jh}^2)}\longleftarrow\frac{1}{2}\sum_{i=1}^N \mu_{q(a_{ijh})}+\alpha_{\psi^2}$
    			        \item[] $\beta_{q(\psi_{jh}^2)}\longleftarrow\beta_{\psi^2}+ \frac{1}{2}\sum_{i=1}^N\mu_{q(a_{ijh})}\big(y_{ij}^2+\mu_{q(\mu_{jh}^2)}+\mu_{q(\lambda_j^2)}\mu_{q(\eta_i^2)}-2y_{ij}\mu_{q(\mu_{jh})}$    \item[]\qquad\qquad\quad$-2y_{ij}\mu_{q(\lambda_j)}\mu_{q(\eta_i)}+2\mu_{q(\mu_{jh})}\mu_{q(\lambda_j)}\mu_{q(\eta_i)}\big)$
    			        \item[] $\mu_{q(\log(\psi_{jh}^2))}\longleftarrow \log\big(\beta_{q(\psi_{jh}^2)}\big)-\mbox{digamma}\big(\alpha_{q(\psi_{jh}^2)}\big)$\,\,;\,\,$\mu_{q(1/\psi_{jh}^2)} \longleftarrow \alpha_{q(\psi_{jh}^2)}\big/\beta_{q(\psi_{jh}^2)}$
                        \item[] If $H>1$: $[\balpha_{q(\bw_j)}]_h\longleftarrow\sum_{i=1}^N\mu_{q(a_{ijh})} + \alpha_w$
                        \end{enumerate}
                        \item[] For $h=1,\ldots,H_j$:
                        \begin{itemize}
						\setlength\itemsep{2pt}
                            \item[] $\mu_{q(\log(w_{jh}))}\longleftarrow\mbox{digamma}\big([\balpha_{q(\bw_j)}]_h\big)-\mbox{digamma}\big(\sum_{h=1}^{H_j}[\balpha_{q(\bw_j)}]_h\big)$
                        \end{itemize}
					\end{itemize}
					\item[] For $i = 1,\ldots,N$:
					\begin{itemize}
						\setlength\itemsep{2pt}
						\item[] $\sigma^2_{q(\eta_i)}\longleftarrow\left(\sum_{j=1}^M\sum_{h=1}^{H_j}\mu_{q(a_{ijh})}\mu_{q(\lambda_j^2)}\mu_{q(1/\psi_{jh}^2)} + \mu_{q(1/\sigma^2)} \right)^{-1}$
    			        \item[]    			      $\mu_{q(\eta_i)}\longleftarrow\sigma^2_{q(\eta_i)}\left\{\mu_{q(1/\sigma^2)}\bx_i^T\bmu_{q(\bbeta)}+\sum_{j=1}^M\sum_{h=1}^{H_j} \mu_{q(a_{ijh})}\mu_{q(\lambda_j)}(y_{ij}-\mu_{q(\mu_{jh})})\mu_{q(1/\psi_{jh}^2)}\right\}$
    			        \item[] $\mu_{q(\eta_i^2)}\longleftarrow\sigma^2_{q(\eta_i)}+\mu_{q(\eta_i)}^2$
					\end{itemize}
                    \item[] $\bSigma_{q(\bbeta)}\longleftarrow \{\mu_{q(1/\sigma^2)}(\sum_{i=1}^N\bx_i\bx_i^T) + \bSigma_{\bbeta}^{-1}\}^{-1}$
                    \item[] $\bmu_{q(\bbeta)}\longleftarrow\bSigma_{q(\bbeta)}\left(\mu_{q(1/\sigma^2)}\sum_{i=1}^N\mu_{q(\eta_i)}\bx_i+\bmu_{\bbeta}\bSigma_{\bbeta}^{-1}\right)$
                    \item[] $\bdeta_{q(\bbeta)}\longleftarrow\left[\begin{array}{cc}
    \bSigma_{q(\bbeta)}^{-1}\bmu_{q(\bbeta)} &
    -\frac{1}{2}\mbox{vec}(\bSigma_{q(\bbeta)}^{-1})
    \end{array}\right]^T$
                    \item[] $\mu_{q(1/\sigma^2)}\longleftarrow \alpha_{q(\sigma^2)}\Big/\left[- \sum_{i=1}^N\left\{G\left(\bdeta_{q(\bbeta)};\bx_i\bx_i^T,\mu_{q(\eta_i)}\bx_i,\mu_{q(\eta_i^2)}\right)\right\}+\beta_{\sigma^2}\right]$
					\end{itemize}
					\item[] \textbf{Relevant Output:} $\alpha_{q(\psi_{jh}^2)}$ $\beta_{q(\psi_{jh}^2)}$, $\mu_{q(\log(w_{jh}))}$, $j = 1,\ldots,M$, $h=1,\ldots,H_j$;  $\mu_{q(\lambda_j)}$, $\sigma^2_{q(\lambda_j)}$, $j = 1,\ldots,M$; $\mu_{q(\eta_i)}$, $\sigma^2_{q(\eta_i)}$, $i = 1,\ldots,N$; $\alpha_{q(\sigma^2)}$, $\beta_{q(\sigma^2)}=\alpha_{q(\sigma^2)}/\mu_{q(1/\sigma^2)}$, $\bmu_{q(\bbeta)}$, $\bSigma_{q(\bbeta)}$.
				\end{itemize}

			\end{small}
		\end{minipage}
	\end{center}
	\caption{\textit{Algorithm for fitting model \eqref{eq:mainSEM} via MFVB.}}
	\label{alg:MFVBmixtureSEMwithLambdaetaAndCovariatesNew}
\end{algorithm}

\FloatBarrier

\subsection{Alternative model}

Non-standard and dynamic patterns in SEMs or latent variable models in general could potentially be handled with model formulations that are different to the one provided in Section \ref{sec:baseModel}. Various works have investigated the use of mixtures of SEMs \citep[e.g.,][]{zhu2001bayesian,vermunt2005structural}; in \cite{asparouhov2016structural} mixtures of SEMs were adopted in conjunction with skewed distributions for the observations and latent variables. The use of an infinite mixture of Dirichlet processes was proposed in \cite{dunson2006bayesian} to characterize the latent response distributions of a factor analytic model nonparametrically.

Here we examine a second SEM formulation where a mixture of Gaussian distribution is imposed to the latent factors and the outcomes are assumed to follow a Gaussian distribution. This alternative formulation resembles the one of \cite{dunson2006bayesian} in a sense that the non-Gaussian features of the data are captured by applying a mixture distribution to the latent variables. The intent here, however is to characterize the effect of distinct subpopulations that may be naturally present (e.g., low and high alcohol drinkers) on the latent factors (cognition in our case), rather than introducing nonparametric components into the model to merely obtain a better model fit. While this model may seem a particularly attractive option to characterize subpopulations of the surveyed individuals, different outcomes can present different multimodal or other non-Gaussian features that can be better captured using model \eqref{eq:mainSEM} instead. Furthermore, fitting a mixture of Gaussians distribution on the latent space may introduce futher identifiability issues related to the estimation of the Gaussian densities parameters and weights. 


The alternative model we propose and study takes the following form for individual $i=1,\ldots,N$, outcome $j=1,\ldots,M$ and mixture component $k=1,\ldots,K$:
\begin{equation}
    \begin{array}{c}
       y_{ij}\,\vert\,\nu_{j},\lambda_j,\eta_i,\psi^2_{j}\sim N(\nu_j+\lambda_j\eta_i,\psi_j^2),\\[1ex] 
       \eta_i\,\vert\,\bbeta_1,\ldots,\bbeta_K,\sigma_1^2,\ldots,\sigma_K^2 \simind\sum_{k=1}^K w_k N(\bx_i^T\bbeta_k,\sigma_k^2)
    \end{array}
\label{eq:altSEM}
\end{equation}
plus priors on $\nu_{j}$, $\lambda_j$, $\psi_j^2$, $\bbeta_k$, $\sigma_k^2$ and $\bw$ such as those used in model \eqref{eq:mainSEM}. As done for model \eqref{eq:mainSEM}, we can set $\lambda_1=1$ and $\lambda_j >0$ for identifiability purposes. In this case identifiability is also a particularly relevant issue affecting the mixture component of the model, which can be solved by applying additional constraints. Similarly to \cite{dunson2006bayesian}, one could set $\nu_1$ to a constant in addition to $\lambda_1$, or use a good initialisation strategy to facilitate convergence, as the one described later on in the real data application.
In the supplementary material we provide a variational inference scheme (Algorithm \ref{alg:MFVBmixtureSEMalt}) for fitting model \eqref{eq:altSEM} via MFVB according to an auxiliary variable representation of the mixture such as the one in \eqref{eq:mainSEM2} and a variational approximating density similar to \eqref{eq:simpSEMcovariatesRestr}. To facilitate convergence and reduce computational times, we recommend fitting a mixture regression to the data (e.g., using the R package `mixtools') and using the estimates of the Gaussian mixture parameters and weights to initialize $\bbeta_k$, $\sigma^2_k$ and $w_k$, for $k=1,\ldots,K$.

\section{Comparing models\label{sec:modelComp}}
\label{subsec:VWAICandVAIC}

A major aspect of model fitting for the proposed SEM formulations is the choice of the numbers of mixture components $H_j$, $j=1,\ldots,M$, for \eqref{eq:mainSEM} and $K$ for \eqref{eq:altSEM}. It is also of interest to identify which model between the baseline formulation \eqref{eq:mainSEM} and the alternative \eqref{eq:altSEM} better fits the data. To this scope, we introduce a variational version of the Watanabe-Akaike information criterion \citep[WAIC;][]{JMLR:v11:watanabe10a,gelman2014understanding} and also consider the variational Akaike information criterion (VAIC) proposed by \cite{you2014variational}. 
Commonly used information criteria typically involve the log posterior density of the observed data given a point estimate and a correction for bias due to overfitting. The point estimate can be the maximum likelihood estimate (as in the Akaike information criterion) or the posterior mean \citep[for the deviance information criterion, DIC;][]{spiegelhalter2002bayesian}, and the bias correction usually corresponds to a measure of the effective number of parameters. 

Unlike DIC, WAIC can be considered a fully Bayesian approach, as its formulation uses the average over the posterior distribution from the model rather than conditioning on a point estimate.
To derive a variational version of WAIC, we start from the following formulation provided by \cite{gelman2014understanding}, which we define for a generic model with a data vector $\by$ of length $N$ and parameter vector $\btheta$: $\text{WAIC} \equiv -2(\text{lppd} - P_{\text{WAIC}})$, with $\text{lppd} \equiv \sum_{i = 1}^N \log \int p(y_i|\btheta)p(\btheta|\by)d\btheta$ representing the logarithmic pointwise predictive density and\\ $P_{\text{WAIC}} \equiv 2\sum_{i=1}^N\left\{ \log (E_{p(\btheta|\by)} p (y_i|\btheta)) - E_{p(\btheta|\by)} (\log p(y_i|\btheta)) \right\}$ being a correction for the effective number of parameters introduced to adjust for overfitting.
%
    %
    %
%
Both lppd and $P_{\text{WAIC}}$ can in practice be obtained using a large number $R$ of samples from the posterior $p(\btheta|\by)$ via MCMC. Noting that $\text{lppd} = \sum_{i = 1}^N \log (E_{p(\btheta|\by)} p(y_i|\btheta))$, a variational version of WAIC immediately arises if the expectation in this expression is computed with respect to the variational density $q(\btheta)$ instead of $p(\btheta|\by)$. We then define the variational WAIC as
\begin{equation}
\begin{array}{c}
    \text{VWAIC} \equiv -2(\text{vlppd} - P_{\text{VWAIC}}),\\[1.5ex]
    \mbox{with}\quad\text{vlppd} \equiv \sum_{i = 1}^N \log( E_{q(\btheta)} p(y_i|\btheta))
    \approx \sum_{i = 1}^N \log \left( \frac{1}{R} \sum_{s=1}^R p(y_i|\btheta^s) \right)\\[1.5ex]
    \quad \begin{array}{cl} \mbox{and}\quad P_{\text{VWAIC}} &\equiv 2\sum_{i=1}^N\left\{ \log (E_{q(\btheta)} p (y_i|\btheta)) - E_{q(\btheta)} (\log p(y_i|\btheta)) \right\}\\[1ex] &\approx 2\sum_{i=1}^N \left\{ \log \left( \frac{1}{R} \sum_{s=1}^R p(y_i|\btheta^s)\right) - \frac{1}{R} \sum_{s=1}^R (\log p(y_i|\btheta^s))\right\},\end{array}
\end{array}
\label{eq:VWAIC}
\end{equation}
and where each $\btheta^s$, $s=1,\ldots,R$, is a vector of the same length as $\btheta$ sampled from $q(\btheta)$. 


The other information criterion we use to compare model fits in the context of variational approximations is VAIC. This criterion can be thought to be an approximation to DIC and is defined as follows: 
\begin{equation}
\begin{array}{c}
    \text{VAIC} \equiv -2(\log p(\by|\tilde{\btheta}) - P_{VAIC}),\\[1.5ex]
    \mbox{with}\quad \tilde{\btheta} \equiv E_{q(\btheta)}(\btheta)\quad\mbox{and}\quad P_{VAIC} \equiv 2\big\{\log p(\by|\tilde{\btheta}) - E_{q(\btheta)} (\log p(\by|\btheta))\big\}.
\end{array}
\label{eq:VAIC}
\end{equation}
In the context of MFVB as presented in this work, $E_q(\btheta)$ can be obtained directly from the approximating densities, for instance using $\mu_{q(\mu_{jh})}$ for the optimal approximating density of $\mu_{jh}$ in \eqref{eq:q_mujh}. As for VWAIC, the expectation $E_{q(\btheta)} (\log p(\by|\btheta))$ can be obtained as an approximation from a large number $R$ of draws from the approximating density $q(\btheta)$. 


\section{Numerical study\label{sec:numStudy}}



\begin{figure}
    \centering
    \includegraphics[width=0.7\linewidth]{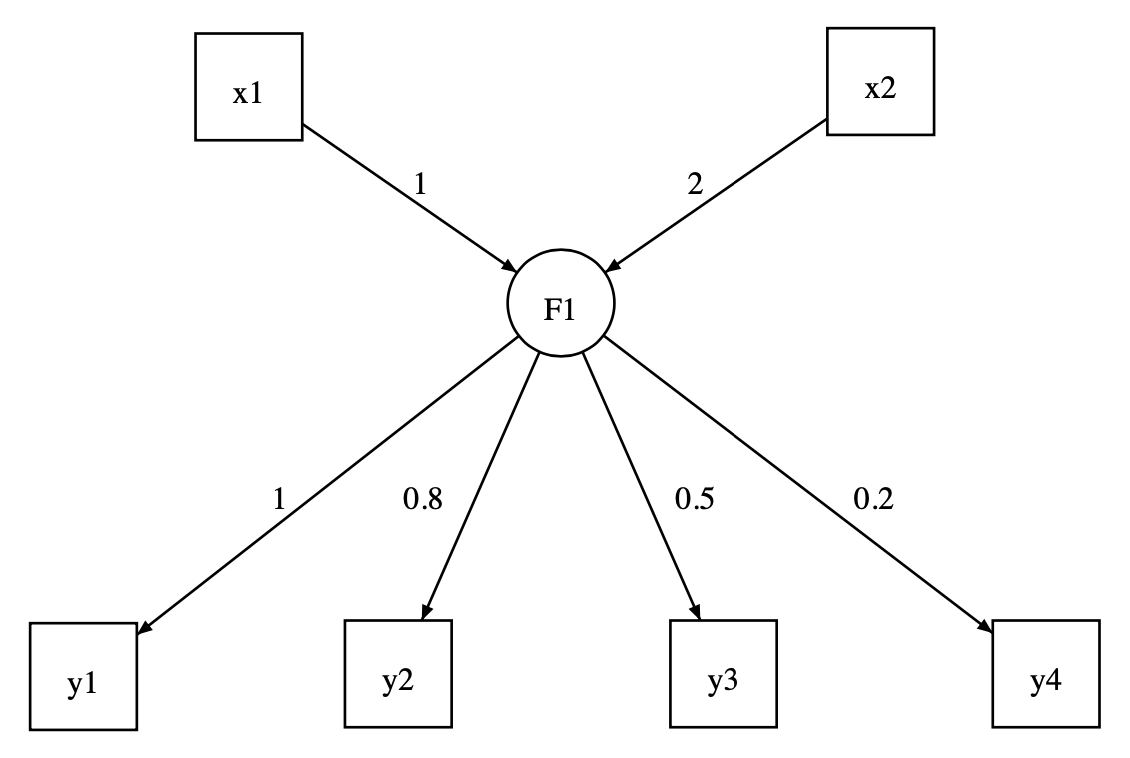}
    \caption{Path diagram for model \eqref{eq:mainSEM} according to the specifications of the simulation example. The numbers on the errors are the true values used to generate the data sets.}
    \label{fig:pathSim}
\end{figure}

The purpose of this section is to study the performance of our MVFB algorithms and the proposed variational information criteria through a simulation exercise. We generated 100 data sets according to model \eqref{eq:mainSEM} with $N=1000$ individuals, $M=4$ outcomes for each individual, and $H_1=1$, $H_2=2$, $H_3=2$ and $H_4=1$ components for the Gaussian mixtures. As a covariate vector we used $\bx_i=(x_{1i},x_{2i})^T$ where the $x_{1i}$'s and $x_{2i}$'s were respectively generated from $N(3,4)$ and $U(0,5)$ distributions, for $i = 1,\dots, N$. We also set $\bbeta=(1,2)^T$, $\lambda_1=1$, $\lambda_2=0.8$, $\lambda_3=0.5$ and $\lambda_4=0.2$ as indicated by the values over the arrows of the Figure \ref{fig:pathSim} path diagram, which shows the relationship between the covariates, the only latent factor and the four outcomes. 
We then ran MFVB using Algorithm \ref{alg:MFVBmixtureSEMwithLambdaetaAndCovariatesNew} and MCMC implemented with the package `rstan' \citep{RStan2024} to fit model \eqref{eq:mainSEM} setting the $H_j$'s to their true values and applying weakly informative priors. Specifically, following the notation in \eqref{eq:mainSEM} we set $\mu_\lambda = 1$, $\sigma^2_\lambda= 1$, $\alpha_{\psi^2} \approx 2.39$, 
$\beta_{\psi^2}\approx 8.69$, 
$\alpha_{\sigma^2} = \beta_{\sigma^2} = 1$, $\mu_{\mu_{jh}} = 0$, $\sigma^2_{\mu_{jh}} = 100$, $\alpha_w = 10$, $\bmu_{\bbeta} = \boldsymbol{0}$ and $\bSigma_{\bbeta}$ to be a $2\times 2$ diagonal matrix with diagonal elements equal to 100. The values of $\alpha_{\psi^2}$ and $\beta_{\psi^2}$ were chosen to impose a prior distribution with mean 6.25 and variance 100. The entire simulation experiment was conducted in R on a personal MacBook Pro laptop with a 2 GHz Quad-Core Intel Core i5 with 16 Gb of memory.

The MFVB and MCMC fits were used to construct 95\% credible intervals for the model parameters to then compute the number of times the true values fell inside the corresponding intervals. These coverage values are presented as percentages in Table \ref{tab:SimCoverage}, while Table \ref{tab:MSE_Sim} presents mean squared errors between the parameter estimates and their true values. For MFVB, the parameter estimates were chosen to be the means of the optimal approximating densities. As an example, the value of $\mu_{q(\mu_{jh})}$ as given in \eqref{eq:q_mujh} and obtained at convergence of the MFVB algorithm was used as an estimate of $\mu_{jh}$. For MCMC, we ran 1 chain of $6,000$ iterations and discarded $3,000$ as burn-in. The MCMC parameter estimates were obtained as means of the remaining $3,000$ samples. 

From Tables \ref{tab:SimCoverage} and \ref{tab:MSE_Sim} it is clear that the mean squared errors of MFVB are comparable to those of MCMC, but the coverage values of MFVB are not very close to the 95\% target. For this reason, we tested the use of nonparametric bootstrap as described in \cite{dang2022fitting} to improve the coverage of MFVB. For each simulated dataset, we obtained 20 bootstrap datasets and ran MFVB on each of them. We then produced percentile bootstrap confidence intervals according to the procedure described in \cite{dang2022fitting} and computed the coverage of these bootstrap intervals for each parameter. The results of MFVB implemented in conjunction with bootrap are stored in Table \ref{tab:SimCoverage} and indicate great improvement over the coverage of plain MFVB, with the coverage values being much closer to the MCMC target. The performance of MFVB could of course be boosted by increasing the number of bootstrap datasets. 

\begin{figure}
    \centering
    \includegraphics[width=0.8\linewidth]{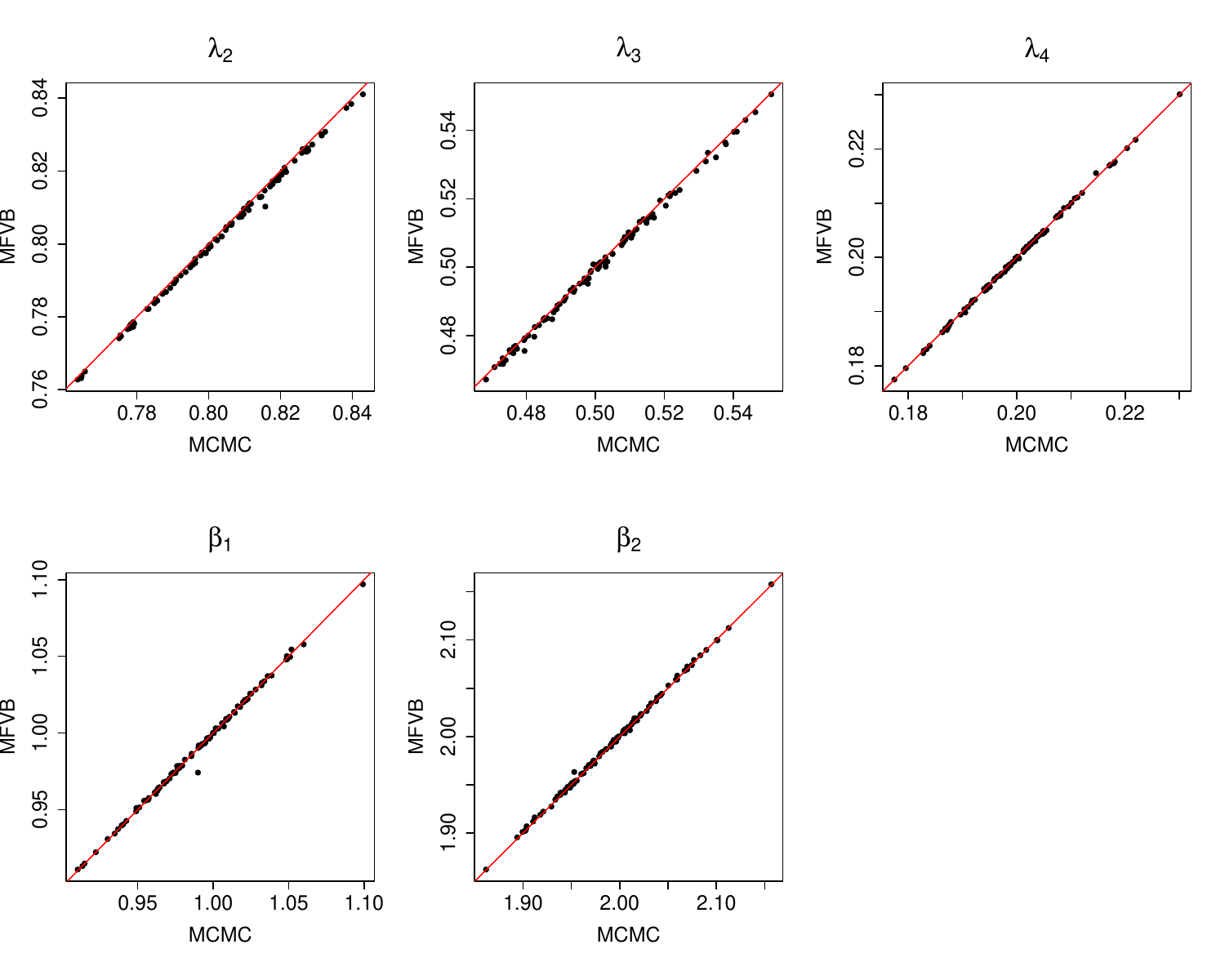}
    \caption{Scatterplots of the mean of the MFVB approximating densities versus the MCMC posterior means for the main parameters of interest (latent variables and covariates coeffiecients). Each point corresponds to a pair of MFVB and MCMC estimates from one of the 100 simulated datasets. The $y=x$ line is represented in red. }
    \label{fig:MFVBvsMCMC}
\end{figure}

We also compared the means of the MFVB approximating densities for $\lambda_2$, $\lambda_3$, $\lambda_4$, $\beta_1$ and $\beta_2$ with the posterior means from MCMC. This comparison is shown in Figure \ref{fig:MFVBvsMCMC} in form of scatterplots for the parameters of interest, with each point corresponding to a pair of MFVB and MCMC estimates from a simulated dataset. The scatterplots show a strong alignment between the MFVB and MCMC estimates.  

\begin{table}
    \centering
    \resizebox{\textwidth}{!}{
    \begin{tabular}{lcccccccccc}
    \hline
      & $\lambda_2$ & $\lambda_3$ &  $\lambda_4$ &$\beta_1 $& $\beta_2$ &$\mu_{11}$ & $\mu_{21}$& $\mu_{22}$ & $\mu_{31}$& $\mu_{32}$\\ \hline
      MFVB & 0.36 &   0.59 &   0.59  &  0.70   & 0.61 &   0.44   & 0.46    &0.60  &  0.50  &  0.84  \\
      MFVB-Bootstrap & 0.86 &   0.92   & 0.83  &  0.89   & 0.84 &   0.85   & 0.84 &   0.78 &   0.92   & 0.91 \\
      MCMC &  0.96 &   0.97  &  0.96  &  0.94 &   0.93   & 0.92   & 0.95  &  0.96   & 0.99  &  0.98 \\[1ex]
      \hline
      & $\mu_{41}$&$\psi^2_{11}$ & $\psi^2_{21}$& $\psi^2_{22}$ & $\psi^2_{31}$& $\psi^2_{32}$ &$\psi^2_{41}$ & $\sigma^2$ \\
      \hline
      MFVB   & 0.58  &  0.86  &  0.76  &  0.89&    0.89  &  0.95  &  0.95   & 0.74   \\
      MFVB- Bootstrap & 0.87  &  0.89  &  0.85 &   0.92  &  0.86  &  0.88  &  0.86  &  0.93 \\
      MCMC & 0.92   & 0.97 &   0.94   & 0.97  &  0.93   & 0.96   & 0.95    &0.98  \\
      \hline
    \end{tabular}}
    \caption{Coverage of MFVB, MFVB with 20 bootstrap iterations and MCMC, computed from fitting model \eqref{eq:mainSEM} as described in the text to 100 simulated datasets.}
    \label{tab:SimCoverage}
\end{table}

\begin{table}
    \centering
    \resizebox{\textwidth}{!}{
    \begin{tabular}{lccccccccc}
    \hline
         & $\lambda_2$ & $\lambda_3$ &  $\lambda_4$ &$\beta_1 $& $\beta_2$ &$\mu_{11}$ & $\mu_{21}$& $\mu_{22}$ & $\mu_{31}$\\ \hline
      MFVB & 0.00033& 0.00037& 0.00009& 0.00130 &0.00293 &0.04802 &0.02825& 0.02854 &0.04056  \\
      MCMC &  0.00034& 0.00038 &0.00009& 0.00131 &0.00297& 0.04865& 0.02799& 0.02843& 0.04001  \\[1ex]
      \hline
      & $\mu_{32}$ & $\mu_{41}$&$\psi^2_{11}$ & $\psi^2_{21}$& $\psi^2_{22}$ & $\psi^2_{31}$& $\psi^2_{32}$ &$\psi^2_{41}$ & $\sigma^2$ \\
      \hline
      MFVB  &0.02911 & 0.00736 &0.05131& 0.04578 &0.02406 &0.12209& 0.06523& 0.00240& 0.02879 \\
      MCMC &0.02848 & 0.00739& 0.05165& 0.03913& 0.02354 &0.12483& 0.06498& 0.00241& 0.02897  \\
      \hline
    \end{tabular}}
    \caption{Mean squared errors of MFVB and MCMC computed from fitting model \eqref{eq:mainSEM} as described in the text to 100 simulated datasets.}
    \label{tab:MSE_Sim}
\end{table}

In addition to the above, we fitted two other models to the same simulated datasets: model \eqref{eq:mainSEM} but with all the $H_j$'s set to 1 so that the outcomes were attributed normal distributions; the alternative model \eqref{eq:altSEM} with $K=2$. For each of these two models and the previously considered model we computed VWAIC and VAIC as defined in \eqref{eq:VWAIC} and \eqref{eq:VAIC} with $R= 1,000$. The first model, i.e. model \eqref{eq:mainSEM} with $H_1=1$, $H_2=2$, $H_3=2$ and $H_4=1$, produced the lowest values of VAIC and VWAIC for all the simulated datasets, meaning that it was always preferable over the other two models in exam. 




\section{Cognitive data\label{sec:application}}

High levels of prenatal alcohol exposure are known to cause fetal alcohol syndrome, a distinct pattern of craniofacial anomalies, growth restriction, and cognitive and behavioural deficits \citep{carter2016fetal,hoyme2005practical,hoyme2016updated,jacobson2004maternal,jacobson2008impaired,mattson2019fetal}. However, some individuals that are subject to prenatal alcohol exposure, especially those at lower levels of exposure, may show cognitive and/or behavioural impairment without exhibiting the characteristic craniofacial dysmorphology. This condition is known as alcohol-related neurodevelopmental disorder. Although the presence of this disorder is often associated with a history of maternal alcohol consumption, the full extent of the dose-response relationship between prenatal alcohol exposure and child cognitive function is yet not well understood. Studying this relationship is crucial in having a proper diagnosis and appropriate treatment and intervention strategy for affected children.

Cognitive function is an ensemble of mental processes that cannot be observed directly and in children or other individuals it is commonly assessed using neuropsychological test results. SEMs are therefore suitable models to relate these measures with the unobserved cognition latent variable. 
\cite{dang2023bsem} used data from studies involving six cohorts in the United States to analyse the effect of prenatal alcohol exposure on child cognition, and their results suggested a strong negative effect of alcohol exposure. The model used by \cite{dang2023bsem} was a multi-cohort SEM which allowed to treat different sets of outcomes across cohorts. In this work we restrict our attention to one of the six studies, specifically the Detroit cohort that is available online at \url{https://doi.org/10.7910/DVN/BOAXR8}. 

In the Detroit study, the mothers were interviewed prenatally or shortly after delivery about their drinking habits during pregnancy, and the children were followed longitudinally to assess their intelligence quotient (IQ), academic achievement in reading and arithmetic, learning and memory abilities (AorLM) and executive function (EF). These cognitive domains were assessed using several tests and the study resulted in 16 measured outcomes from 377 children. The first 3 outcomes are related to IQ tests, the 4th to 10th outcomes are related to EF and the remaining 6 are measures of AorLM. We name these outcomes according to the description provided by the online data source, as shown in the Figure \ref{fig:pathDetroit} path diagram. For example, ``IQ 1'' and ``EF 3'' stand for the outcomes of the first IQ test and the third EF test.
The mothers' consumption of beer, wine and liquor during pregnancy were summarised in terms of ounces of absolute alcohol consumed per day (1 ounce absolute alcohol equals 2 standard drinks of alcohol), which is denoted as ``aadxp".  Following \cite{dang2023bsem}, the covariates used to study the data are $\log(\text{aadxp}+1)$ and a pre-computed propensity score that accounts for other confounders (see \citeauthor{akkaya2021propensity}, \citeyear{akkaya2021propensity}, for more details about this propensity score).

The model used by \cite{dang2023bsem} to study each cohort assumed that the outcomes are jointly normal given the latent variable, and therefore was unable to fully capture non-Gaussian features such as skewness and multimodality of some challenging outcomes. In such a situation, model \eqref{eq:mainSEM} is particularly suitable to study the data since it allows to model each outcome using mixtures of Gaussians with varying numbers of components. Alternatively, one can assume that the non-Gaussian features are due to the presence of distinct groups with distinct drinking habits (for instance, a group of people less affected by alcohol exposure, and one strongly affected) and assume a Gaussian mixture on the latent variable space as in model \eqref{eq:altSEM}, although this implies the actual presence of a grouped structure in the data. In this section we consider both model formulations. 

\subsection{Fitted models and priors}

One of the advantages of formulating SEMs in a Bayesian framework is that it is possible to incorporate prior knowledge about the parameters via the prior distribution. Given the relatively large number of observed outcomes, $16$, and small number of participants, $377$, with many test results missing for several participants (the median percentage of missing data across outcomes is $22.944\%$), choosing priors adequately plays an important role also in facilitating identifiability and convergence of Bayesian inference algorithms. Following previous analyses of the same data \citep{dang2023bsem,jacobson2021effects}, the observations corresponding to each outcome were standardized to have standard deviation of $15$, which is approximately that of IQ.

Referring to the priors of model \eqref{eq:mainSEM} and their hyperparameters, we set our priors as follows. For the latent factor parameters that are not constrained to $1$ for identifiability we set $\mu_\lambda=1$ and $\sigma^2_\lambda=1$. For the mixture component-specific intercepts we use $\sigma^2_{\mu_{jh}}=10^4$ and for each $j$ we set the $\mu_{\mu_{jh}}$'s to the estimated intercepts of a mixture of regressions of the $j$th outcome against the covariates obtained through the function `regmixEM' from the package `mixtools' if $H_j>1$, or the intercept of a simple regression of the $j$th outcome against the covariates when $H_j=1$. Based on the estimates of the regressions used to initialize the $\mu_{\mu_{jh}}$'s and the fact that the outcomes were standardised to have standard deviation equal to $15$, we set $\alpha_{\psi^2}=\alpha_{\sigma^2}=6$ and $\beta_{\psi^2} = \beta_{\sigma^2}=500$ so that the Inverse Gamma priors on $\psi_{jh}^2$ and $\sigma^2$ had means and variances respectively equal to $100$ and $2,500$. To impose a weakly informative prior on $w_{jh}$, we set $\alpha_w=10$. Finally, we set $\bmu_\beta=\boldsymbol{0}$ and $\bSigma_\beta$ to be a diagonal matrix with diagonal elements equal to 100.

%


We also fitted the alternative model \eqref{eq:altSEM} using $K= 2$ as the number of components of the Gaussian mixture on the latent variable and a Dirichlet prior with $\alpha_w=5$ for the components weights, i.e. a smaller value of $\alpha_w$ compared to the one used for model \eqref{eq:mainSEM} as the fitting of the Gaussian mixture was more challenging on the latent space than on the outcomes. The priors and corresponding hyperparameters for $\bbeta$, $\lambda_j$, $\psi_j$ and $\sigma^2_k$ were kept the same as the ones used for model \eqref{eq:mainSEM}. We used a $N(0,10^4)$ prior for each $\nu_j$.

We considered and fitted four models as follows: 1) model \eqref{eq:mainSEM} choosing the number of components from visual inspection of the empirical density functions of the outcomes, which resulted in $(H_1,\ldots,H_{16})=(1, 1, 2, 1, 2, 2, 2, 1, 1, 1, 2, 2, 1, 2, 1, 1)$ (option 1); 2) model \eqref{eq:mainSEM} with all the $H_j$'s set to 1, i.e. with the mixtures collapsing to Gaussian densities (option 2); 3) model \eqref{eq:mainSEM} choosing the number of components through the function \texttt{densityMclust} from the R package `mclust', which resulted in $(H_1,\ldots,H_{16})=(1, 1, 2, 1, 2, 2, 2, 1, 1, 3, 2, 2, 1, 2, 1, 1)$ (option 3); 4) model \eqref{eq:altSEM} with a mixture of 2 components on the latent factor (option 4).

\subsection{Results}

We fitted the four options listed above via MFVB using Algorithms \ref{alg:MFVBmixtureSEMwithLambdaetaAndCovariatesNew} and \ref{alg:MFVBmixtureSEMalt}, and for each model computed VAIC and VWAIC as described in Section \ref{subsec:VWAICandVAIC} by approximating the expectation $E_{q(\btheta)} (\log p(\by|\btheta))$ in both \eqref{eq:VWAIC} and \eqref{eq:VAIC} using $R = 10,000$ random draws from the variational approximating densities. The values collected in Table \ref{tab:vaic} indicate that option 1, i.e. model \eqref{eq:mainSEM} with the number of components chosen from visual inspection of the outcomes distributions, is the preferred option. Option 3 gave similar values of the information criteria to option 1 and this is due to the fact that the only difference between the two options is a different number of components for the 10$^{th}$ outcome, with $H_{10}$ being equal to $1$ in the case of option 1 and $3$ for option 3. 

It is interesting to note from Table \ref{tab:betaModel1to3} that the estimates of the regression coefficients of options 1, 2, and 3 (which all refer to model \eqref{eq:mainSEM} fittings) are very similar, meaning that fitting a Gaussian mixture on the outcomes can improve the predictive power without significantly affecting inference. Similarly, Table \ref{tab:betaModel4} reporting the estimates of $\bbeta_1$, $\bbeta_2$ and $w_k$ from model \eqref{eq:altSEM} considered in option 4 suggests that the effects of the two covariates averaged over the two Gaussian mixture components are close to those from options 1, 2 and 3. 
We also note the similarity between the results presented here and the pooled effect size estimate of prenatal alcohol exposure shown in Table 4 of \cite{AkkayaHocagil2022Meta}, where the small discrepancy may be due to some differences in the set of outcomes being used.

\begin{table}
    \centering
    \begin{tabular}{ccccc}
    \hline
         & Option 1  & Option 2 & Option 3 & Option 4 \\ \hline
     VAIC    & \textbf{213641.5} &  221791.1& 214995 & 221783.7\\
     VWAIC & \textbf{137814.1}& 139496.5 & 138604.3 & 138367.5 \\ \hline
    \end{tabular}
    \caption{VWAIC and VAIC of the models fitted in each setting, based on 10,000 samples generated from the variational approximating densities.}
    \label{tab:vaic}
\end{table}

\begin{figure}
    \centering
    \includegraphics[width=0.9\linewidth]{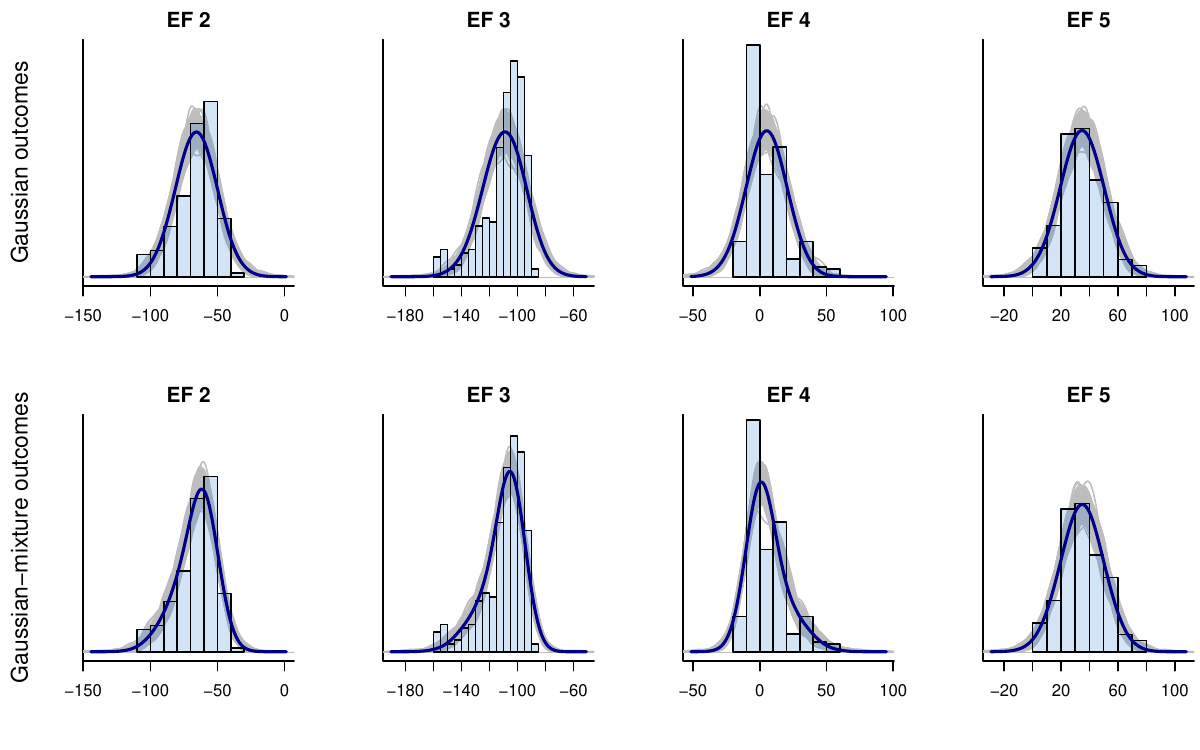}
    \caption{Posterior predictive densities for four representative outcomes of the Detroit data plotted over histograms of the outcome measurements. The top row refers to the model where the conditional densities of the outcomes are all assumed to be Gaussian according to option 2, and the bottom row corresponds to model \eqref{eq:mainSEM} fitted according to option 1. Each grey line is the kernel density estimate of one of the $300$ drawings of $\by$ from the posterior predictive distribution; their averages are represented as blue lines.}
    \label{fig:DetroitPPCs}
\end{figure}

Another way to check whether a model fits the data well is by looking at posterior predictive densities. Recall that the posterior predictive distribution of a new data point $\tilde{y}$ given the observed data $\by$ is 
$p(\tilde{y}|\by) = \int p(\tilde{y}|\btheta)p(\btheta|\by)d\btheta$.
In most cases it is not possible to compute $p(\tilde{y}|\by)$ in an analytic form and a simple way to visualize this distribution is by generating new data sets with values of $\btheta$ drawn from the posterior. These simulated data can be plotted together with the observed data to assess the quality of the model fit. We performed such checks using the models of options 1 and 2. However, instead of drawing $\btheta$ from the posterior $p(\btheta|\by)$, we generated data from the approximating density $q(\btheta)$ as specified in Section 2.1 with the optimal parameters obtained from Algorithm \ref{alg:MFVBmixtureSEMwithLambdaetaAndCovariatesNew} at convergence.  Figure \ref{fig:DetroitPPCs} shows the kernel density estimates of $300$ drawings from the posterior predictive distribution displayed over histograms of the measurements of a selection of four outcomes that showed non-Gaussian features. From the plots it is evident that for the last outcome, whose histogram looks roughly bell shaped, the two models fit the data equally well. However, when the data show non-Gaussian features such as skewness as for the remaining outcomes, the model of option 1 captures the shape of the data better than that of option 2. Similar plots for all outcomes and models are provided in the supplementary material \ref{sec:additionalResultApplication}.


\begin{table}
    \centering
    \begin{tabular}{cccc}
    \hline
         & Option 1  & Option 2 & Option 3 \\ \hline
      $\hat{\beta}_1$   & -0.36646 & -0.44148 & -0.38889  \\
      & (0.61319) & (0.59784) & (0.61857)\\
      $\hat{\beta}_2$   & -6.89255 & -7.04235 & -6.97791  \\
      & (1.93116) & (1.88431) & (1.94755)\\
      \hline
    \end{tabular}
    \caption{Estimates of the elements of $\bbeta$ and their standard deviations (in brackets) for the models of options 1, 2 and 3 obtained as means and standard deviations from the variational approximating density of $\bbeta$.}
    \label{tab:betaModel1to3}
\end{table}

\begin{table}
    \centering
    \begin{tabular}{ccccc}
    \hline
       \multicolumn{2}{c}{1st component } & & \multicolumn{2}{c}{2nd component} \\ \hline
       $\hat{\beta}_{1,1}$  & 0.29766 & & $\hat{\beta}_{2,1}$ & -1.07275 \\
       & (0.80827) & & & (0.85658) \\
       $\hat{\beta}_{2,1}$  &-4.59911 & & $\hat{\beta}_{2,2}$ &  -8.46553 \\ 
       & (2.54153) & & & (2.63300) \\
       $\hat{w}_1$ & 0.48368 & & $\hat{w}_2$ & 0.51632\\ \hline
    \end{tabular}
    \caption{Estimates of the elements of $\bbeta_1$, $\bbeta_2$ and their standard deviations (in brackets) for the two mixture components of model \eqref{eq:altSEM} according to option 4 obtained as means and standard deviations from the variational approximating densities of $\bbeta_1$ and $\bbeta_2$. The estimates of the weights are also provided and these coincide with the means of the corresponding approximating densities. }
    \label{tab:betaModel4}
\end{table}

\section{Conclusions\label{sec:conclusion}}
We have introduced two SEM formulations that make use of mixtures of Gaussian densities to better capture non-Gaussian features of the data and take into account missing data. We have also proposed a new variational information criterion that can be computed from the output of MFVB algorithms at convergence, VWAIC, to compare candidate models. Through a simulation study, we have shown that both VWAIC and the alternative VAIC we have also examined are able to identify the correct model.
We have fitted the models to data from a study on prenatal alcohol exposure and child cognition in Detroit using MFVB and shown that the model with Gaussian mixtures on the outcomes is suitable for capturing the non-Gaussian features of the data.  

Our real data application is also an example of incorporating exogenous covariates in a SEM, although it has to be noted that the small number of participants with a high level of alcohol exposure in the Detroit study could hinder estimation of the relationship between alcohol exposure and cognition. Considering data from multiple studies or cohorts would allow to fit more flexible structures, including those with non-linear or semi-parametric regression components, and obtain more reliable estimates of the effect of prenatal alcohol exposure on cognition. The variational framework we have presented here could easily be extended to such multi-group settings, but we leave this for future investigation. 


\section{Acknowledgements}
Khue-Dung Dang was supported by the Andrew Sisson support package from the School of Mathematics and Statistics at the University of Melbourne. 

\bibliographystyle{apalike}
\bibliography{ref}

\null\vfill\eject
%
\appendix
\renewcommand{\thealgorithm}{S.\arabic{algorithm}}
\setcounter{algorithm}{0}
\renewcommand{\thesection}{S.\arabic{section}}
\setcounter{section}{0}
\renewcommand{\theequation}{S.\arabic{equation}}
\renewcommand{\thesection}{S.\arabic{section}}
\renewcommand{\thetable}{S.\arabic{table}}
\renewcommand{\thefigure}{S.\arabic{figure}}
\setcounter{equation}{0}
\setcounter{table}{0}
\setcounter{section}{0}
\setcounter{figure}{0}
\setcounter{page}{1}
\setcounter{footnote}{0}

\centerline{\Large Supplement for:}
\vskip5mm
\centerline{\Large\bf Variational Bayes for Mixture of Gaussian}
\vskip4mm
\centerline{\Large\bf Structural Equation Models}
\vskip7mm
\ifthenelse{\boolean{UnBlinded}}{
\centerline{\normalsize\sc By Khue-Dung Dang$\null^{1}$\footnotemark, Luca Maestrini$\null^{2}$ and Francis K.C. Hui$\null^{2}$}
\vskip5mm
\centerline{\textit{$\null^1$University of Melbourne and $\null^2$Australian National University}}}{}

\section{Derivations for model with Gaussian mixture on the responses}

The joint likelihood function of model \eqref{eq:mainSEM} is 
\begin{align*}
    p(\by,\bmu,\blambda,\bdeta,\bpsi^2,\bbeta,\sigma^2,\ba,\bw) =
    & \prod_{(i,j)\in\mathcal{S}_{\tiny\mbox{obs}}}  \prod_{h=1}^{H_j} \left[(2\pi\psi_{jh}^2)^{-1/2} \exp \left\{-  \frac{(y_{ij}-\mu_{jh}-\lambda_j\eta_i)^2}{2\psi_{jh}^2}\right\}\right]^{a_{ijh}}\\
    \times&\prod_{i=1}^N\left[(2\pi\sigma^2)^{-1/2}\exp\left\{-\frac{(\eta_i-\bx_i^T\bbeta)^2}{2\sigma^2}\right\}\right]\\
    \times&\prod_{j=2}^M\left[(2\pi\sigma^2_\lambda)^{-1/2}\exp\left\{-\frac{(\lambda_j-\mu_\lambda)^2}{2\sigma^2_\lambda}\right\}\right]\\
    \times&\prod_{j=1}^M\prod_{h=1}^{H_j}\Bigg(\left[(2\pi\sigma^2_{\mu_{jh}})^{-1/2}\exp\left\{-\frac{(\mu_{jh}-\mu_{\mu_{jh}})^2}{2\sigma^2_{\mu_{jh}}}\right\}\right]\\
    &\qquad\qquad\times\left[\left(\beta_{\psi^2}\right)^{\alpha_{\psi^2}}\left\{\Gamma\left(\alpha_{\psi^2}\right)\right\}^{-1}(\psi_{jh}^2)^{-\alpha_{\psi^2}-1}\exp\left(-\frac{\beta_{\psi^2}}{\psi_{jh}^2}\right)\right]\Bigg)\\
    \times&(2\pi)^{-p/2}\vert\bSigma_{\bbeta}\vert^{-1/2}\exp\left\{-\frac{1}{2}(\bbeta-\bmu_{\bbeta})^T\bSigma_{\bbeta}^{-1}(\bbeta-\bmu_{\bbeta})\right\}\\
    \times&\left(\beta_{\sigma^2}\right)^{\alpha_{\sigma^2}}\left\{\Gamma\left(\alpha_{\sigma^2}\right)\right\}^{-1}(\sigma^2)^{-\alpha_{\sigma^2}-1}\exp\left(-\frac{\beta_{\sigma^2}}{\sigma^2}\right)\\
    \times & \prod_{(i,j)\in\mathcal{S}_{\tiny\mbox{obs}}} \frac{1}{a_{ij1}!\ldots a_{ijH_j}!}w_{j1}^{a_{ij1}}\ldots w_{jH_j}^{a_{ijH_j}}\\
    \times & \prod_{j=1}^M\left\{\frac{\Gamma(H_j\alpha_w)}{\Gamma(\alpha_w)^{H_j}}\prod_{h=1}^{H_j}w_{jh}^{\alpha_w-1}\right\}.
\end{align*}
The logarithm of the joint likelihood function arising from model \eqref{eq:mainSEM} is
\begin{align*}
    &\log p(\by;\bmu,\bdeta,\blambda,\bpsi^2,\bbeta,\sigma^2,\ba,\bw)= \sum_{(i,j)\in\mathcal{S}_{\tiny\mbox{obs}}} \sum_{h=1}^{H_j}  \left\{- \frac{1}{2}\log(2\pi) - \frac{(y_{ij}-\mu_{jh}-\lambda_j\eta_i)^2}{2\psi_{jh}^2}  \right\}a_{ijh}\\
    &-\frac{N+2\alpha_{\sigma^2}+2}{2}\log(\sigma^2)-\frac{1}{2\sigma^2}\left[\sum_{i=1}^N\Big\{\eta_i^2-2\eta_i\bx_i^T\bbeta+(\bx_i^T\bbeta)^2\Big\}+2\beta_{\sigma^2}\right] \\
    &-\frac{1}{2}\sum_{(i,j)\in\mathcal{S}_{\tiny\mbox{obs}}}\sum_{h=1}^{H_j}a_{ijh}\log(\psi_{jh}^2)-\sum_{j=1}^M\sum_{h=1}^{H_j}\left(\alpha_{\psi^2} +1\right)\log(\psi_{jh}^2)-\beta_{\psi^2}\sum_{j=1}^M\sum_{h=1}^{H_j}\frac{1}{\psi_{jh}^2}\\
    &-\frac{1}{2\sigma^2_\lambda}\sum_{j=2}^M\lambda_j^2+\frac{\mu_\lambda}{\sigma^2_\lambda}\sum_{j=2}^M\lambda_j-\frac{1}{2\sigma^2_{\mu_{jh}}}\sum_{j=1}^M\sum_{h=1}^{H_j}\mu_{jh}^2+\frac{\mu_{\mu_{jh}}}{\sigma^2_{\mu_{jh}}}\sum_{j=1}^M\sum_{h=1}^{H_j}\mu_{jh}-\frac{1}{2}(\bbeta-\bmu_{\bbeta})^T\bSigma_{\bbeta}^{-1}(\bbeta-\bmu_{\bbeta}) \\
    &+ \sum_{(i,j)\in\mathcal{S}_{\tiny\mbox{obs}}}\sum_{h=1}^{H_j}  \{a_{ijh}\log(w_{jh})-\log(a_{ijh}!)\}+(\alpha_w-1)\sum_{j=1}^M\sum_{h=1}^{H_j}\log(w_{jh})+\mbox{const}.
\end{align*}
In order to achieve a tractable MFVB approximation for model \eqref{eq:mainSEM}, we factorize the density approximating the posterior as in equation \eqref{eq:simpSEMcovariatesRestr}.

For $j=1,\ldots,M$ and $h=1,\ldots,H_j$,
\begin{equation*}
    p(\mu_{jh}\,\vert\,\mbox{rest})\propto\exp\left[-\frac{\mu_{jh}^2}{2}\left(\sum_{i:(i,j)\in\mathcal{S}_{\tiny\mbox{obs}}}\frac{a_{ijh}}{\psi_{jh}^2}+\frac{1}{\sigma^2_{\mu_{jh}}}\right)+\mu_{jh}\left\{\sum_{i:(i,j)\in\mathcal{S}_{\tiny\mbox{obs}}}\frac{a_{ijh}(y_{ij}-\lambda_j\eta_i)}{\psi_{jh}^2}+\frac{\mu_{\mu_{jh}}}{\sigma^2_{\mu_{jh}}}\right\}\right],
\end{equation*}
from which it follows that
\begin{equation*}
\begin{array}{c}
    q^*(\mu_{jh})\quad\mbox{is}\quad N\big(\mu_{q(\mu_{jh})},\sigma^2_{q(\mu_{jh})}\big),\quad j=1,\ldots,M,\,\,h=1,\ldots,H_j,\\[3ex]
    \mbox{with}\quad\mu_{q(\mu_{jh})}=\sigma^2_{q(\mu_{jh})}\left\{\sum_{i:(i,j)\in\mathcal{S}_{\tiny\mbox{obs}}}\mu_{q(a_{ijh})}\mu_{q(1/\psi_{jh}^2)}\big(y_{ij}-\mu_{q(\lambda_j)}\mu_{q(\eta_i)}\big)+\dfrac{\mu_{\mu_{jh}}}{\sigma^2_{\mu_{jh}}}\right\}\\[3ex]
    \mbox{and}\quad\sigma^2_{q(\mu_{jh})}=\left(\sum_{i:(i,j)\in\mathcal{S}_{\tiny\mbox{obs}}}\mu_{q(a_{ijh})}\mu_{q(1/\psi_{jh}^2)} + \dfrac{1}{\sigma^2_{\mu_{jh}}} \right)^{-1}.
\end{array}
\end{equation*}

Without loss of generality, we fix $\lambda_1$ to 1 for ensuring identifiability. For $j=2,\ldots,M$,
\begin{align*}
 p(\lambda_j\,\vert\,\mbox{rest})&\propto\exp\Bigg[ -\frac{\lambda_j^2}{2} \Bigg( \sum_{i:(i,j)\in\mathcal{S}_{\tiny\mbox{obs}}}\sum_{h=1}^{H_j} \frac{a_{ijh}\eta_i^2}{\psi_{jh}^2}+\frac{1}{\sigma^2_\lambda} \Bigg)\\[2ex] 
 &\qquad\qquad+ \lambda_j \Bigg\{\sum_{i:(i,j)\in\mathcal{S}_{\tiny\mbox{obs}}}\sum_{h=1}^{H_j}a_{ijh}\eta_i\left(\frac{y_{ij}-\mu_{jh}}{\psi_{jh}^2} \right)+\frac{\mu_{\lambda}}{\sigma^2_{\lambda}}\Bigg\}\Bigg],
\end{align*}
from which it follows that 
\begin{equation*}
\begin{array}{c}
    q^*(\lambda_j)\quad\mbox{is}\quad N\big(\mu_{q(\lambda_j)},\sigma^2_{q(\lambda_j)}\big),\quad j=2,\ldots,M,\\[3ex]
    \mbox{with}\quad\mu_{q(\lambda_j)}=\sigma^2_{q(\lambda_j)}\left\{\sum_{i:(i,j)\in\mathcal{S}_{\tiny\mbox{obs}}}\sum_{h=1}^{H_j}\mu_{q(a_{ijh})}\mu_{q(\eta_i)}\mu_{q(1/\psi_{jh}^2)}\big(y_{ij} -\mu_{q(\mu_{jh})}\big)+\dfrac{\mu_{\lambda}}{\sigma^2_{\lambda}}\right\}\\[3ex]
    \mbox{and}\quad\sigma^2_{q(\lambda_j)}=\left(\sum_{i:(i,j)\in\mathcal{S}_{\tiny\mbox{obs}}}\sum_{h=1}^{H_j} \mu_{q(a_{ijh})}\mu_{q(\eta_i^2)}\mu_{q(1/\psi_{jh}^2)}+\dfrac{1}{\sigma^2_\lambda} \right)^{-1}.
\end{array}
\end{equation*}

For $i=1,\ldots,N$,
\begin{align*}
    p(\eta_i\,\vert\,\mbox{rest})&\propto\exp\Bigg[-\frac{\eta_i^2}{2} \Bigg(\sum_{j:(i,j)\in\mathcal{S}_{\tiny\mbox{obs}}}\sum_{h=1}^{H_j}\frac{a_{ijh}\lambda_j^2}{\psi_{jh}^2} + \frac{1}{\sigma^2}\Bigg)\\[2ex]
    &\qquad\qquad+ \eta_i \Bigg\{\sum_{j:(i,j)\in\mathcal{S}_{\tiny\mbox{obs}}}\sum_{h=1}^{H_j} a_{ijh}\lambda_j\Bigg( \frac{y_{ij}-\mu_{jh}}{\psi_{jh}^2}\Bigg) + \frac{\bx_i^T\bbeta}{\sigma^2}\Bigg\} \Bigg],
\end{align*}
from which it follows that
\begin{equation*}
\begin{array}{c}
    q^*(\eta_i)\quad\mbox{is}\quad N\big(\mu_{q(\eta_i)},\sigma^2_{q(\eta_i)}\big),\quad 1=1,\ldots,N,\\[3ex]
    \mbox{with}\quad\mu_{q(\eta_i)}=\sigma^2_{q(\eta_i)}\left\{\mu_{q(1/\sigma^2)}\bx_i^T\bmu_{q(\bbeta)}+\sum_{j:(i,j)\in\mathcal{S}_{\tiny\mbox{obs}}}\sum_{h=1}^{H_j} \mu_{q(a_{ijh})}\mu_{q(\lambda_j)}(y_{ij}-\mu_{q(\mu_{jh})})\mu_{q(1/\psi_{jh}^2)}\right\}\\[3ex]
    \mbox{and}\quad\sigma^2_{q(\eta_i)}=\left(\sum_{j:(i,j)\in\mathcal{S}_{\tiny\mbox{obs}}}\sum_{h=1}^{H_j}\mu_{q(a_{ijh})}\mu_{q(\lambda_j^2)}\mu_{q(1/\psi_{jh}^2)} + \mu_{q(1/\sigma^2)} \right)^{-1}.
\end{array}
\end{equation*}

For $j=1,\ldots,M$ and $h=1,\ldots,H_j$,
\begin{equation*}
    p(\psi_{jh}^2\,\vert\,\mbox{rest})\propto(\psi_{jh}^2)^{-\left(\sum_{i:(i,j)\in\mathcal{S}_{\tiny\mbox{obs}}} \frac{a_{ijh}}{2}+\alpha_{\psi^2} +1\right)}\exp\Bigg[ -\frac{1}{\psi_{jh}^2}\Bigg\{\beta_{\psi^2}+\sum_{i:(i,j)\in\mathcal{S}_{\tiny\mbox{obs}}}\frac{a_{ijh}}{2}(y_{ij}-\mu_{jh}-\lambda_j\eta_i)^2 \Bigg\}\Bigg],
\end{equation*}
from which it follows that
\begin{equation*}
\begin{array}{c}
    q^*(\psi_{jh}^2)\quad\mbox{is}\quad \mbox{Inverse-Gamma}\Big(\alpha_{q(\psi_{jh}^2)},\beta_{q(\psi_{jh}^2)}\Big),\quad j=1,\ldots,M,\,\,h=1,\ldots,H_j,\\[3ex]
    \mbox{with}\quad\alpha_{q(\psi_{jh}^2)}\equiv\dfrac{1}{2}\sum_{i:(i,j)\in\mathcal{S}_{\tiny\mbox{obs}}} \mu_{q(a_{ijh})}+\alpha_{\psi^2}\\[3ex]
    \mbox{and}\quad\beta_{q(\psi_{jh}^2)}\equiv\beta_{\psi^2}+ \dfrac{1}{2}\sum_{i:(i,j)\in\mathcal{S}_{\tiny\mbox{obs}}}\mu_{q(a_{ijh})}\big(y_{ij}^2+\mu_{q(\mu_{jh}^2)}+\mu_{q(\lambda_j^2)}\mu_{q(\eta_i^2)}-2y_{ij}\mu_{q(\mu_{jh})}\\[1ex]    -2y_{ij}\mu_{q(\lambda_j)}\mu_{q(\eta_i)}+2\mu_{q(\mu_{jh})}\mu_{q(\lambda_j)}\mu_{q(\eta_i)}\big).
\end{array}
\end{equation*}

Next,
\begin{align*}
    p(\sigma^2\,\vert\,\mbox{rest})&\propto(\sigma^2)^{-(N+2\alpha_{\sigma^2}+2)/2}\exp\left(-\frac{1}{2\sigma^2}\left[\sum_{i=1}^N\left\{\eta_i^2-2\eta_i\bx_i^T\bbeta+(\bx_i^T\bbeta)^2\right\}+2\beta_{\sigma^2}\right]\right).
\end{align*}
Following \cite{wand2017fast}, we define the function
\begin{align*}
    G\left(\left[\begin{array}{c}
    \bv_1\\
    \bv_2
    \end{array}\right];\bQ,\br,s\right)&\equiv-\frac{1}{8}\tr\left(\bQ\{\vecof^{-1}(\bv_2)\}^{-1}[\bv_1\bv_1^T\{\vecof^{-1}(\bv_2)\}^{-1}-2\bI]\right)\\[1ex]
    &\quad-\frac{1}{2}\br^T\{\vecof^{-1}(\bv_2)\}^{-1}\bv_1-\frac{1}{2}s.
\end{align*}
for a $d\times 1$ vector $\bv_1$ and a $d^2\times 1$ vector $\bv_2$ such that $\vecof^{-1}(\bv_2)$ is symmetric.
This function arises from the fact that if $\btheta$ is a $d\times 1$ Multivariate Normal random vector with natural parameter vector $\bdeta$, then
\begin{equation*}
    E_{\btheta}\left\{-\frac{1}{2}\left(\btheta^T\bQ\btheta-2\br^T\btheta+s\right)\right\}=E_{\btheta}\left(-\frac{1}{2}\btheta^T\bQ\btheta+\br^T\btheta\right)-\frac{1}{2}s=G(\bdeta;\bQ,\br,s).
\end{equation*}
Given the expression of $p(\sigma^2\,\vert\,\mbox{rest})$ given above, we want to calculate
\begin{equation*}
    E_{\bmu_{q(\bbeta)}}\left[-\frac{1}{2}\left\{\mu_{q(\eta_i^2)}-2\mu_{q(\eta_i)}\bx_i^T\bbeta+(\bx_i^T\bbeta)^2\right\}\right]=E_{\bmu_{q(\bbeta)}}\left[-\frac{1}{2}\left\{(\bbeta^T\bx_i\bx_i^T\bbeta-2\mu_{q(\eta_i)}\bx_i^T\bbeta+\mu_{q(\eta_i^2)}\right\}\right],
\end{equation*}
which is related to the expectation involving the $G$ function with $\btheta=\bbeta$, $\bQ=\bx_i\bx_i^T$, $\br=\mu_{q(\eta_i)}\bx_i$ and $s=\mu_{q(\eta_i^2)}$. It means this involves calculation of $G\left(\bdeta_{q(\bbeta)};\bx_i\bx_i^T,\mu_{q(\eta_i)}\bx_i,\mu_{q(\eta_i^2)}\right)$, where $\bdeta_{q(\bbeta)}$ is defined later.
The optimal approximating density arising from $p(\sigma^2\,\vert\,\mbox{rest})$ is then
\begin{equation*}
\begin{array}{c}
    q^*(\sigma^2)\quad\mbox{is}\quad\mbox{Inverse-Gamma}\left(\alpha_{q(\sigma^2)},\beta_{q(\sigma^2)}\right)\\[2ex]
    \mbox{with}\quad\alpha_{q(\sigma^2)}\equiv\dfrac{N+2\alpha_{\sigma^2}}{2}\\[2ex]
    \mbox{and}\quad\beta_{q(\sigma^2)}\equiv - \sum_{i=1}^N\left\{G\left(\bdeta_{q(\bbeta)};\bx_i\bx_i^T,\mu_{q(\eta_i)}\bx_i,\mu_{q(\eta_i^2)}\right)\right\}+\beta_{\sigma^2}.
\end{array}
\end{equation*}

Using the property $\mbox{tr}(\bA^T\bB)=\mbox{vec}(\bB)^T\mbox{vec}(\bA)$ and noticing that $\bx_i^T\bbeta\bbeta^T\bx_i=\mbox{tr}(\bx_i^T\bbeta\bbeta^T\bx_i)=\mbox{tr}(\bx_i\bx_i^T\bbeta\bbeta^T)=\mbox{vec}(\bbeta\bbeta^T)^T\mbox{vec}(\bx_i\bx_i^T)$ and $\bbeta^T\bSigma_{\bbeta}^{-1}\bbeta=\mbox{tr}(\bbeta^T\bSigma_{\bbeta}^{-1}\bbeta)=\mbox{tr}(\bSigma_{\bbeta}^{-1}\bbeta\bbeta^T)=\mbox{vec}(\bbeta\bbeta^T)^T\mbox{vec}(\bSigma_{\bbeta}^{-1})$
we have
\begin{align*}
    p(\bbeta\,\vert\,\mbox{rest})&\propto\exp\left\{-\frac{1}{2\sigma^2}\sum_{i=1}^N\left(\bx_i^T\bbeta\bbeta^T\bx_i-2\eta_i\bx_i^T\bbeta\right)-\frac{1}{2}\bbeta^T\bSigma_{\bbeta}^{-1}\bbeta+\bbeta^T\bSigma_{\bbeta}^{-1}\bmu_{\bbeta}\right\}\\[2ex]
    &=\exp\left(\left[\begin{array}{c}
    \bbeta\\[1.5ex]
    \mbox{vec}(\bbeta\bbeta^T)
    \end{array}\right]^T\left[\begin{array}{c}
    \dfrac{1}{\sigma^2}\sum_{i=1}^N\eta_i\bx_i+\bSigma_{\bbeta}^{-1}\bmu_{\bbeta}\\[1.5ex]
    -\dfrac{1}{2\sigma^2}\mbox{vec}(\sum_{i=1}^N\bx_i\bx_i^T) -\dfrac{1}{2}\mbox{vec}(\bSigma_{\bbeta}^{-1})
    \end{array}\right]\right).
\end{align*}
Hence, using (S.4) of \cite{wand2017fast},
\begin{equation*}
\begin{array}{c}
    q^*(\bbeta)\quad\mbox{is}\quad N\left(\bmu_{q(\bbeta)},\bSigma_{q(\bbeta)}\right)\\[2ex]  
    \mbox{with}\quad\bmu_{q(\bbeta)}\equiv\bSigma_{q(\bbeta)}\left(\mu_{q(1/\sigma^2)}\sum_{i=1}^N\mu_{q(\eta_i)}\bx_i+\bSigma_{\bbeta}^{-1}\bmu_{\bbeta}\right)    \\[2ex]
    \mbox{and}\quad\bSigma_{q(\bbeta)}\equiv \left\{\mu_{q(1/\sigma^2)}\left(\sum_{i=1}^N\bx_i\bx_i^T\right) + \bSigma_{\bbeta}^{-1}\right\}^{-1},
\end{array}
\end{equation*}
from which we define
\begin{equation*}
    \bdeta_{q(\bbeta)}\equiv\left[\begin{array}{c}
    \bSigma_{q(\bbeta)}^{-1}\bmu_{q(\bbeta)}\\[1ex]
    -\frac{1}{2}\mbox{vec}(\bSigma_{q(\bbeta)}^{-1})
    \end{array}\right].
\end{equation*}

For $(i,j)\in\mathcal{S}_{\tiny\mbox{obs}}$, 
\begin{align*}
    p(\ba_{ij}\,\vert\,\mbox{rest})  \propto\prod_{h = 1}^{H_j} \frac{1}{a_{ijh}!} \Bigg[w_{jh}(\psi_{jh}^2)^{-1/2}(2\pi)^{-1/2} \exp \Bigg\{-  \frac{(y_{ij}-\mu_{jh}-\lambda_j\eta_i)^2}{2\psi_{jh}^2}\Bigg\}  \Bigg]^{a_{ijh}},
\end{align*}
from which it follows that 
\begin{equation*}
\begin{array}{c}
    q^*(\ba_{ij})\quad\mbox{is}\quad\mbox{Multinomial}\big(1;\mu_{q(\ba_{ij})}\big),\quad i=1,\ldots,N,\quad k=1,\ldots, K,\\[2ex]
    \mbox{with}\quad\mu_{q(a_{ijh})}\equiv\exp(\tau_{ijh})\big/\sum_{h=1}^{H_j}\exp(\tau_{ijh}),\quad h=1,\ldots,H_j,\\[3ex]
    \mbox{and}\quad\tau_{ijh}\equiv \mu_{q(\log(w_{jh}))}-\dfrac{1}{2}\mu_{q(\log(\psi_{jh}^2))}-\dfrac{1}{2}\log(2\pi)\\[2ex] -\dfrac{1}{2}\mu_{q(1/\psi_{jh}^2)}\big(y_{ij}^2+\mu_{q(\mu^2_{jh})}+\mu_{q(\lambda_j^2)}\mu_{q(\eta_i^2)}-2y_{ij}\mu_{q(\mu_{jh})}-2y_{ij}\mu_{q(\lambda_j)}\mu_{q(\eta_i)}+2\mu_{q(\mu_{jh})}\mu_{q(\lambda_j)}\mu_{q(\eta_i)}\big).
\end{array}
\end{equation*}

Note that
\begin{equation*}
  p(\bw_j\,\vert\,\mbox{rest})\propto \prod_{h=1}^{H_j}w_{jh}^{\big(\sum_{(i,j)\in\mathcal{S}_{\tiny\mbox{obs}}}{a_{ijh}} + \alpha_w-1\big)},\quad j=1,\ldots,M,
\end{equation*}
from which it follows that
\begin{equation*}
\begin{array}{c}
    q^*(\bw_j)\quad\mbox{is}\quad \mbox{Dirichlet}(H_j,\balpha_{q(\bw_j)}),\\[2ex]
    \mbox{with}\quad\balpha_{q(\bw_j)}=\big(\sum_{(i,j)\in\mathcal{S}_{\tiny\mbox{obs}}}\mu_{q(a_{ij1})} + \alpha_w,\ldots,\sum_{(i,j)\in\mathcal{S}_{\tiny\mbox{obs}}}\mu_{q(a_{ijH_j})} + \alpha_w\big).
\end{array}
\end{equation*}
This provides $\mu_{q(\log(w_{jh}))}=\mbox{digamma}\big([\balpha_{q(\bw_j)}]_h\big)-\mbox{digamma}\big(\sum_{h=1}^{H_j}[\balpha_{q(\bw_j)}]_h\big)$.

\section{Derivations for model with Gaussian mixture on latent factors}

%
       %
       %
        %
        %
        %
%
Another way to represent model \eqref{eq:altSEM} is, for $i=1,\ldots,N$, $j=1,\ldots,M$ and $k=1,\ldots,K$,
\begin{equation}
    \begin{array}{c}
        y_{ij}\,\vert\,\nu_{j},\lambda_j,\eta_i,\psi^2_{j}\sim N(\nu_j+\lambda_j\eta_i,\psi_j^2),\\[1ex]
        p(\eta_i\,\vert\,\bbeta_1,\ldots,\bbeta_K,\sigma_1^2,\ldots,\sigma_K^2,\ba_i)= \prod_{k=1}^{K} \left[\sigma_k^{-1}(2\pi)^{-1/2} \exp \left\{-  \dfrac{(\eta_i-\bx_i^T\bbeta_k)^2}{2\sigma_k^2}\right\}\right] ^{a_{ik}}, \\[1.5ex]
        \ba_i \simind \mbox{Multinomial}(1; \bw),\,\,\,\nu_{j}\simind N(\mu_{\nu},\sigma^2_{\nu}),\,\,\,\lambda_j\simind N(\mu_{\lambda},\sigma^2_\lambda),\,\,\,\psi_j^2\simind\mbox{Inverse-Gamma}(\alpha_{\psi^2},\beta_{\psi^2}),\\[1ex] 
        \bbeta_k\sim N(\bmu_{\bbeta},\bSigma_{\bbeta}),\quad\sigma_k^2\sim\mbox{Inverse-Gamma}(\alpha_{\sigma^2},\beta_{\sigma^2}),\quad 
        \bw\sim\mbox{Dirichlet}(K,\alpha_w).
    \end{array}
\label{eq:altSEMbis}
\end{equation}
The joint likelihood function of model \eqref{eq:altSEMbis} is
\begin{align*}
    &p(\by,\bnu,\blambda,\bdeta,\psi_1^2,\ldots,\psi_M^2,\bbeta_1,\ldots,\bbeta_K,\sigma_1^2,\ldots,\sigma^2_K,\ba,\bw)\\
    =&\prod_{(i,j)\in\mathcal{S}_{\tiny\mbox{obs}}}  (2\pi\psi_j^2)^{-1/2} \exp \left\{-  \frac{(y_{ij}-\nu_j-\lambda_j\eta_i)^2}{2\psi_j^2}\right\}\\
    \times&\prod_{i=1}^N\prod_{k=1}^{K} \left[\sigma_k^{-1}(2\pi)^{-1/2} \exp \left\{-  \dfrac{(\eta_i-\bx_i^T\bbeta_k)^2}{2\sigma_k^2}\right\}\right] ^{a_{ik}}\\
    \times&\prod_{j=2}^M\left[(2\pi\sigma^2_\lambda)^{-1/2}\exp\left\{-\frac{(\lambda_j-\mu_\lambda)^2}{2\sigma^2_\lambda}\right\}\right]\\
    \times&\prod_{j=1}^M\Bigg(\left[(2\pi\sigma^2_\nu)^{-1/2}\exp\left\{-\frac{(\nu_j-\mu_\nu)^2}{2\sigma^2_\nu}\right\}\right]\\
    &\qquad\qquad\times\left[\left(\beta_{\psi^2}\right)^{\alpha_{\psi^2}}\left\{\Gamma\left(\alpha_{\psi^2}\right)\right\}^{-1}(\psi_j^2)^{-\alpha_{\psi^2}-1}\exp\left(-\frac{\beta_{\psi^2}}{\psi_j^2}\right)\right]\Bigg)\\
    \times&\prod_{k=1}^K(2\pi)^{-p/2}\vert\bSigma_{\bbeta}\vert^{-1/2}\exp\left\{-\frac{1}{2}(\bbeta_k-\bmu_{\bbeta})^T\bSigma_{\bbeta}^{-1}(\bbeta_k-\bmu_{\bbeta})\right\}\\
    \times&\prod_{k=1}^K\left(\beta_{\sigma^2}\right)^{\alpha_{\sigma^2}}\left\{\Gamma\left(\alpha_{\sigma^2}\right)\right\}^{-1}(\sigma_k^2)^{-\alpha_{\sigma^2}-1}\exp\left(-\frac{\beta_{\sigma^2}}{\sigma_k^2}\right)\\
    \times & \prod_{i=1}^N \frac{1}{a_{i1}!\ldots a_{iK}!}w_1^{a_{i1}}\ldots w_K^{a_{iK}}\\
    \times & \frac{\Gamma(K\alpha_w)}{\Gamma(\alpha_w)^K}\prod_{k=1}^K w_k^{\alpha_w-1}.
\end{align*}
The logarithm of the joint likelihood function arising from model \eqref{eq:altSEM} is
\begin{align*}
    &\log p(\by;\bnu,\blambda,\bdeta,\psi_1^2,\ldots,\psi_M^2,\bbeta_1,\ldots,\bbeta_K,\sigma_1^2,\ldots,\sigma^2_K,\ba,\bw)\\
    =&-\sum_{(i,j)\in\mathcal{S}_{\tiny\mbox{obs}}} \frac{(y_{ij}-\nu_j-\lambda_j\eta_i)^2}{2\psi_j^2}\\
    &- \frac{\sum_{i=1}^N a_{ik}+2\alpha_{\sigma^2}+2}{2}\sum_{k =1}^K\log(\sigma_k^2)-\frac{1}{2}\sum_{k=1}^K \frac{1}{\sigma_k^2}\left[\sum_{i=1}^Na_{ik}\Big\{\eta_i^2-2\eta_i\bx_i^T\bbeta_k+(\bx_i^T\bbeta_k)^2\Big\}+2\beta_{\sigma^2}\right] \\
    &-\frac{1}{2}\sum_{(i,j)\in\mathcal{S}_{\tiny\mbox{obs}}}\log(\psi_j^2)-\sum_{j=1}^M\left(\alpha_{\psi^2} +1\right)\log(\psi_j^2)-\beta_{\psi^2}\sum_{j=1}^M\frac{1}{\psi_j^2}\\
    &-\frac{1}{2\sigma^2_\lambda}\sum_{j=2}^M\lambda_j^2+\frac{\mu_\lambda}{\sigma^2_\lambda}\sum_{j=2}^M\lambda_j-\frac{1}{2\sigma^2_\nu}\sum_{j=1}^M\nu_j^2+\frac{\mu_\nu}{\sigma^2_\nu}\sum_{j=1}^M\nu_j\\
    &-\frac{1}{2}\sum_{k=1}^K(\bbeta_k-\bmu_{\bbeta})^T\bSigma_{\bbeta}^{-1}(\bbeta_k-\bmu_{\bbeta}) \\
    &+ \sum_{i=1}^N  \{a_{ik}\log(w_k)-\log(a_{ik}!)\}+(\alpha_w-1)\sum_{k=1}^K\log(w_k)+\mbox{const}.
\end{align*}
In order to achieve a tractable MFVB approximation for model \eqref{eq:altSEM}, we factorize the density approximating the posterior as follows:
\begin{align}
\begin{split}
    &q(\bnu,\blambda,\bdeta,\psi_1^2,\ldots,\psi_M^2,\bbeta_1,\ldots,\bbeta_K,\sigma_1^2,\ldots,\sigma^2_K,\ba,\bw)\\
    &=q(\bw)\prod_{i=1}^N\left\{q(\eta_i)q(\ba_i)\right\}\prod_{j=1}^M \left\{q(\nu_j)q(\psi_j^2)\right\}\prod_{j=2}^M q(\lambda_j)\prod_{k=1}^K\left\{q(\bbeta_k)q(\sigma^2_k)\right\}.
\end{split}
    \label{eq:altSEMcovariatesRestr}
\end{align}

For $j=1,\ldots,M$,
\begin{equation*}
    p(\nu_j\,\vert\,\mbox{rest})\propto\exp\left\{-\frac{\nu_j^2}{2}\left(\sum_{i:(i,j)\in\mathcal{S}_{\tiny\mbox{obs}}}\frac{1}{\psi_j^2}+\frac{1}{\sigma^2_{\nu}}\right)+\nu_j\left(\sum_{i:(i,j)\in\mathcal{S}_{\tiny\mbox{obs}}}\frac{y_{ij}-\lambda_j\eta_i}{\psi_j^2}+\frac{\mu_{\nu}}{\sigma^2_{\nu}}\right)\right\},
\end{equation*}
from which it follows that
\begin{equation*}
\begin{array}{c}
    q^*(\nu_j)\quad\mbox{is}\quad N\big(\mu_{q(\nu_j)},\sigma^2_{q(\nu_j)}\big),\quad j=1,\ldots,M,\\[3ex]
    \mbox{with}\quad\mu_{q(\nu_j)}=\sigma^2_{q(\nu_j)}\left\{\sum_{i:(i,j)\in\mathcal{S}_{\tiny\mbox{obs}}}\mu_{q(1/\psi_j^2)}\big(y_{ij}-\mu_{q(\lambda_j)}\mu_{q(\eta_i)}\big)+\dfrac{\mu_{\nu}}{\sigma^2_{\nu}}\right\}\\[3ex]
    \mbox{and}\quad\sigma^2_{q(\nu_j)}=\left(\sum_{i:(i,j)\in\mathcal{S}_{\tiny\mbox{obs}}}\mu_{q(1/\psi_j^2)} + \dfrac{1}{\sigma^2_{\nu}} \right)^{-1}.
\end{array}
\end{equation*}

Without loss of generality, we fix $\lambda_1$ to 1 for ensuring identifiability. For $j=2,\ldots,M$,
\begin{equation*}
 p(\lambda_j\,\vert\,\mbox{rest})\propto\exp\Bigg[ -\frac{\lambda_j^2}{2} \Bigg( \sum_{i:(i,j)\in\mathcal{S}_{\tiny\mbox{obs}}}\frac{\eta_i^2}{\psi_j^2}+\frac{1}{\sigma^2_\lambda} \Bigg)+ \lambda_j \Bigg\{\sum_{i:(i,j)\in\mathcal{S}_{\tiny\mbox{obs}}}\eta_i\left(\frac{y_{ij}-\nu_j}{\psi_j^2} \right)+\frac{\mu_{\lambda}}{\sigma^2_{\lambda}}\Bigg\}\Bigg],
\end{equation*}
from which it follows that
\begin{equation*}
\begin{array}{c}
    q^*(\lambda_j)\quad\mbox{is}\quad N\big(\mu_{q(\lambda_j)},\sigma^2_{q(\lambda_j)}\big),\quad j=2,\ldots,M,\\[3ex]
    \mbox{with}\quad\mu_{q(\lambda_j)}=\sigma^2_{q(\lambda_j)}\left\{\sum_{i:(i,j)\in\mathcal{S}_{\tiny\mbox{obs}}}\mu_{q(\eta_i)}\mu_{q(1/\psi_j^2)}\big(y_{ij} -\mu_{q(\nu_j)}\big)+\dfrac{\mu_{\lambda}}{\sigma^2_{\lambda}}\right\}\\[3ex]
    \mbox{and}\quad\sigma^2_{q(\lambda_j)}=\left(\sum_{i:(i,j)\in\mathcal{S}_{\tiny\mbox{obs}}} \mu_{q(\eta_i^2)}\mu_{q(1/\psi_j^2)}+\dfrac{1}{\sigma^2_\lambda} \right)^{-1}.
\end{array}
\end{equation*}

For $i=1,\ldots,N$,
\begin{align*}
    p(\eta_i\,\vert\,\mbox{rest})&\propto\exp\Bigg[-\frac{\eta_i^2}{2} \Bigg(\sum_{j:(i,j)\in\mathcal{S}_{\tiny\mbox{obs}}}\frac{\lambda_j^2}{\psi_j^2} + \sum_{k=1}^K\frac{a_{ik}}{\sigma_k^2}\Bigg)\\[2ex]
    &\qquad\qquad+ \eta_i \Bigg\{\sum_{j:(i,j)\in\mathcal{S}_{\tiny\mbox{obs}}}\lambda_j\Bigg( \frac{y_{ij}-\nu_j}{\psi_j^2}\Bigg) + \sum_{k=1}^K\frac{a_{ik}\bx_i^T\bbeta_k}{\sigma_k^2}\Bigg\} \Bigg],
\end{align*}
from which it follows that
\begin{equation*}
\begin{array}{c}
    q^*(\eta_i)\quad\mbox{is}\quad N\big(\mu_{q(\eta_i)},\sigma^2_{q(\eta_i)}\big),\quad i=1,\ldots,N,\\[3ex]
    \mbox{with}\quad\mu_{q(\eta_i)}=\sigma^2_{q(\eta_i)}\left\{\sum_{k=1}^K\mu_{q(a_{ik})}\mu_{q(1/\sigma_k^2)}\bx_i^T\bmu_{q(\bbeta_k)}+\sum_{j:(i,j)\in\mathcal{S}_{\tiny\mbox{obs}}}\mu_{q(\lambda_j)}(y_{ij}-\mu_{q(\nu_j)})\mu_{q(1/\psi_j^2)}\right\}\\[3ex]
    \mbox{and}\quad\sigma^2_{q(\eta_i)}=\left(\sum_{j:(i,j)\in\mathcal{S}_{\tiny\mbox{obs}}}\mu_{q(\lambda_j^2)}\mu_{q(1/\psi_j^2)} + \sum_{k=1}^K\mu_{q(a_{ik})}\mu_{q(1/\sigma_k^2)} \right)^{-1}.
\end{array}
\end{equation*}

For $j=1,\ldots,M$,
\begin{equation*}
    p(\psi_j^2\,\vert\,\mbox{rest})\propto(\psi_j^2)^{-\left(\sum_{i:(i,j)\in\mathcal{S}_{\tiny\mbox{obs}}} \frac{1}{2}+\alpha_{\psi^2} +1\right)}\exp\Bigg[ -\frac{1}{\psi_j^2}\Bigg\{\beta_{\psi^2}+\sum_{i:(i,j)\in\mathcal{S}_{\tiny\mbox{obs}}}\frac{1}{2}(y_{ij}-\nu_j-\lambda_j\eta_i)^2 \Bigg\}\Bigg],
\end{equation*}
from which it follows that
\begin{equation*}
\begin{array}{c}
    q^*(\psi_j^2)\quad\mbox{is}\quad \mbox{Inverse-Gamma}\Big(\alpha_{q(\psi_j^2)},\beta_{q(\psi_j^2)}\Big),\quad j=1,\ldots,M,\\[3ex]
    \mbox{with}\quad\alpha_{q(\psi_j^2)}\equiv\sum_{i:(i,j)\in\mathcal{S}_{\tiny\mbox{obs}}} \frac{1}{2}+\alpha_{\psi^2}\\[3ex]
    \mbox{and}\quad\beta_{q(\psi_j^2)}\equiv\beta_{\psi^2}+ \dfrac{1}{2}\sum_{i:(i,j)\in\mathcal{S}_{\tiny\mbox{obs}}}\big(y_{ij}^2+\mu_{q(\nu_j^2)}+\mu_{q(\lambda_j^2)}\mu_{q(\eta_i^2)}-2y_{ij}\mu_{q(\nu_j)}\\[1ex]    -2y_{ij}\mu_{q(\lambda_j)}\mu_{q(\eta_i)}+2\mu_{q(\nu_j)}\mu_{q(\lambda_j)}\mu_{q(\eta_i)}\big).
\end{array}
\end{equation*}

Next, for $k=1,\ldots,K$,
\begin{align*}
    p(\sigma_k^2\,\vert\,\mbox{rest})&\propto(\sigma_k^2)^{-(\sum_{i=1}^N a_{ik}+2\alpha_{\sigma^2}+2)/2}\exp\left(-\frac{1}{2\sigma_k^2}\left[\sum_{i=1}^N a_{ik}\left\{\eta_i^2-2\eta_i\bx_i^T\bbeta_k+(\bx_i^T\bbeta_k)^2\right\}+2\beta_{\sigma^2}\right]\right).
\end{align*}
Given this expression, we want to calculate
\begin{align*}
    &E_{\bmu_{q(\bbeta_k)}}\left[-\frac{1}{2}\left\{\mu_{q(\eta_i^2)}-2\mu_{q(\eta_i)}\bx_i^T\bbeta_k+(\bx_i^T\bbeta_k)^2\right\}\right]\\[1ex]
    &\qquad\qquad=E_{\bmu_{q(\bbeta_k)}}\left[-\frac{1}{2}\left\{(\bbeta_k^T\bx_i\bx_i^T\bbeta_k-2\mu_{q(\eta_i)}\bx_i^T\bbeta_k+\mu_{q(\eta_i^2)}\right\}\right],
\end{align*}
which is related to the expectation involving the $G$ function with $\btheta=\bbeta_k$, $\bQ=\bx_i\bx_i^T$, $\br=\mu_{q(\eta_i)}\bx_i$ and $s=\mu_{q(\eta_i^2)}$. It means we have to calculate $G\left(\bdeta_{q(\bbeta_k)};\bx_i\bx_i^T,\mu_{q(\eta_i)}\bx_i,\mu_{q(\eta_i^2)}\right)$, where $\bdeta_{q(\bbeta_k)}$ is defined later.
The optimal approximating density arising from $p(\sigma_k^2\,\vert\,\mbox{rest})$ is then
\begin{equation*}
\begin{array}{c}
    q^*(\sigma_k^2)\quad\mbox{is}\quad\mbox{Inverse-Gamma}\left(\alpha_{q(\sigma_k^2)},\beta_{q(\sigma_k^2)}\right),\quad k=1,\ldots, K,\\[2ex]
    \mbox{with}\quad\alpha_{q(\sigma_k^2)}\equiv\dfrac{\sum_{i=1}^N \mu_{q(a_{ik})}+2\alpha_{\sigma^2}}{2}\\[2ex]
    \mbox{and}\quad\beta_{q(\sigma_k^2)}\equiv - \sum_{i=1}^N\mu_{q(a_{ik})}\left\{G\left(\bdeta_{q(\bbeta_k)};\bx_i\bx_i^T,\mu_{q(\eta_i)}\bx_i,\mu_{q(\eta_i^2)}\right)\right\}+\beta_{\sigma^2}.
\end{array}
\end{equation*}

For $k=1,\ldots,K$,
\begin{align*}
    p(\bbeta_k\,\vert\,\mbox{rest})&\propto\exp\left\{-\frac{1}{2\sigma_k^2}\sum_{i=1}^N a_{ik}\left(\bx_i^T\bbeta_k\bbeta_k^T\bx_i-2\eta_i\bx_i^T\bbeta_k\right)-\frac{1}{2}\bbeta_k^T\bSigma_{\bbeta}^{-1}\bbeta_k+\bbeta_k^T\bSigma_{\bbeta}^{-1}\bmu_{\bbeta}\right\}\\[2ex]
    &=\exp\left(\left[\begin{array}{c}
    \bbeta_k\\[1.5ex]
    \mbox{vec}(\bbeta_k\bbeta_k^T)
    \end{array}\right]^T\left[\begin{array}{c}
    \dfrac{1}{\sigma_k^2}\sum_{i=1}^N a_{ik}\eta_i\bx_i+\bSigma_{\bbeta}^{-1}\bmu_{\bbeta}\\[1.5ex]
    -\dfrac{1}{2\sigma_k^2}\mbox{vec}(\sum_{i=1}^N a_{ik}\bx_i\bx_i^T) -\dfrac{1}{2}\mbox{vec}(\bSigma_{\bbeta}^{-1})
    \end{array}\right]\right).
\end{align*}
Hence, using (S.4) or Wand (2017, JASA),
\begin{equation*}
\begin{array}{c}
    q^*(\bbeta_k)\quad\mbox{is}\quad N\left(\bmu_{q(\bbeta_k)},\bSigma_{q(\bbeta_k)}\right),\quad k=1,\ldots,K,\\[2ex]  
    \mbox{with}\quad\bmu_{q(\bbeta_k)}\equiv\bSigma_{q(\bbeta_k)}\left(\mu_{q(1/\sigma_k^2)}\sum_{i=1}^N\mu_{q(a_{ik})}\mu_{q(\eta_i)}\bx_i+\bSigma_{\bbeta}^{-1}\bmu_{\bbeta}\right)    \\[2ex]
    \mbox{and}\quad\bSigma_{q(\bbeta_k)}\equiv \left\{\mu_{q(1/\sigma_k^2)}\left(\sum_{i=1}^N\mu_{q(a_{ik})}\bx_i\bx_i^T\right) + \bSigma_{\bbeta}^{-1}\right\}^{-1},
\end{array}
\end{equation*}
from which we define
\begin{equation*}
    \bdeta_{q(\bbeta_k)}\equiv\left[\begin{array}{c}
    \bSigma_{q(\bbeta_k)}^{-1}\bmu_{q(\bbeta_k)}\\[1ex]
    -\frac{1}{2}\mbox{vec}(\bSigma_{q(\bbeta_k)}^{-1})
    \end{array}\right].
\end{equation*}

For $i=1,\ldots,N$ and $k=1,\ldots,K$, 
\begin{align*}
    p(\ba_i\,\vert\,\mbox{rest})  \propto\prod_{k=1}^K\frac{1}{a_{ik}!} \Bigg[w_k(\sigma_k^2)^{-1/2}(2\pi)^{-1/2}\exp\Bigg\{-\frac{\eta_i^2-2\eta_i\bx_i^T\bbeta_k+(\bx_i^T\bbeta_k)^2}{2\sigma_k^2}\Bigg\}\Bigg]^{a_{ik}},
\end{align*}
from which it follows that
\begin{equation*}
\begin{array}{c}
    q^*(\ba_i)\quad\mbox{is}\quad\mbox{Multinomial}\big(1;\mu_{q(\ba_i)}\big),\quad i=1,\ldots,N,\\[2ex]
    \mbox{with}\quad\mu_{q(a_{ik})}\equiv\exp(\tau_{ik})\big/\sum_{k=1}^K\exp(\tau_{ik}),\quad k=1,\ldots, K\\[3ex]
    \mbox{and}\quad\tau_{ik}\equiv \mu_{q(\log(w_k))}-\dfrac{1}{2}\mu_{q(\log(\sigma_k^2))}-\dfrac{1}{2}\log(2\pi)+\mu_{q(1/\sigma_k^2)}\left\{G\left(\bdeta_{q(\bbeta_k)};\bx_i\bx_i^T,\mu_{q(\eta_i)}\bx_i,\mu_{q(\eta_i^2)}\right)\right\}.
\end{array}
\end{equation*}
Note that
\begin{equation*}
  p(\bw\,\vert\,\mbox{rest})\propto \prod_{k=1}^{K}w_k^{\big(\sum_{i=1}^N a_{ik} + \alpha_w-1\big)},\quad k=1,\ldots,K, 
\end{equation*}
from which it follows that
\begin{equation*}
\begin{array}{c}
    q^*(\bw)\quad\mbox{is}\quad \mbox{Dirichlet}(K,\balpha_{q(\bw)}),\\[2ex]
    \mbox{with}\quad\balpha_{q(\bw)}=\big(\sum_{i=1}^N\mu_{q(a_{i1})} + \alpha_w,\ldots,\sum_{i=1}^N\mu_{q(a_{iK})} + \alpha_w\big).
\end{array}
\end{equation*}
This provides $\mu_{q(\log(w_{k}))}=\mbox{digamma}\big([\balpha_{q(\bw)}]_k\big)-\mbox{digamma}\big(\sum_{k=1}^{K}[\balpha_{q(\bw)}]_k\big)$.

All the above derivations produce Algorithm \ref{alg:MFVBmixtureSEMalt} for fitting model \eqref{eq:altSEMbis}.

\begin{algorithm}[!th]
	\begin{center}
		\begin{minipage}[t]{154mm}
			\begin{small}
				\begin{itemize}
					\setlength\itemsep{2pt}
					\item[] \textbf{Data Input:} $\by_i$, $i=1,\ldots,N$, vectors of length $M$.
					\item[] \textbf{Hyperparameter Input:} $\mu_\nu ,\mu_{\lambda}\in\mathbb{R}$, $\sigma_\nu, \sigma_\lambda,\alpha_{\psi^2},\beta_{\psi^2},\alpha_{\sigma^2},\beta_{\sigma^2}\in\mathbb{R}^+$, $\alpha_w=1$, $\bmu_{\bbeta}\in\mathbb{R}^p$, $\bSigma_{\bbeta}$ being a $p\times p$ symmetric positive definite matrix.
					\item[] \textbf{Initialize:}  $\mu_{q(\nu_j)} \in \mathbb{R}$, $\mu_{q(1/\psi_j^2)}\in\mathbb{R}^+$, $j=1,\ldots,M$; $\mu_{q(\lambda_j)}\in \mathbb{R},\mu_{q(\lambda_j^2)}\in \mathbb{R}^+$, $j=2,\ldots,M$; $\mu_{q(\eta_i)}\in \mathbb{R},\mu_{q(\eta_i^2)}\in \mathbb{R}^+$, $i=1,\ldots,N$; $\mu_{q(1/\sigma_k^2)}\in\mathbb{R}^+$, $\mu_{q(\log(w_k))}\in \mathbb{R}^{-}$,  $k=1,\ldots,K$; $\mu_{q(\lambda_1)}=\mu_{q(\lambda_1^2)}=1$; $\mu_{q(a_{ik})}\in(0,1)$, $i=1,\ldots,N$, $k=1,\ldots,K$; $\alpha_{q(\psi_j^2)}\longleftarrow\sum_{i:(i,j)\in\mathcal{S}_{\tiny\mbox{obs}}} \frac{1}{2}+\alpha_{\psi^2}$, $j=1,\ldots,M$.
					\item[] \textbf{Cycle until convergence:}
                    \begin{itemize}
						\setlength\itemsep{2pt}
                            \item[] For $k=1,\ldots,K$:
                            \begin{itemize}
                            \setlength\itemsep{2pt}
                                \item[] $\bSigma_{q(\bbeta_k)}\longleftarrow \left\{\mu_{q(1/\sigma_k^2)}\left(\sum_{i=1}^N\mu_{q(a_{ik})}\bx_i\bx_i^T\right) + \bSigma_{\bbeta}^{-1}\right\}^{-1}$
                                \item[] $\bmu_{q(\bbeta_k)}\longleftarrow\bSigma_{q(\bbeta_k)}\left(\mu_{q(1/\sigma_k^2)}\sum_{i=1}^N\mu_{q(a_{ik})}\mu_{q(\eta_i)}\bx_i+\bSigma_{\bbeta}^{-1}\bmu_{\bbeta}\right)$
                                \item[] $\bdeta_{q(\bbeta_k)}\longleftarrow\left[\begin{array}{cc}
                                \bSigma_{q(\bbeta_k)}^{-1}\bmu_{q(\bbeta_k)} &
                                -\frac{1}{2}\mbox{vec}(\bSigma_{q(\bbeta_k)}^{-1})
                                \end{array}\right]^T$
                                \item[] $\alpha_{q(\sigma_k^2)}\longleftarrow\frac{1}{2}\sum_{i=1}^N \mu_{q(a_{ik})}+\alpha_{\sigma^2}$
                                \item[] $\beta_{q(\sigma_k^2)}\longleftarrow -  \sum_{i=1}^N\mu_{q(a_{ik})}\left\{G\left(\bdeta_{q(\bbeta_k)};\bx_i\bx_i^T,\mu_{q(\eta_i)}\bx_i,\mu_{q(\eta_i^2)}\right)\right\}+\beta_{\sigma^2}$
                                \item[] $\mu_{q(\log(\sigma_k^2))}\longleftarrow \log\big(\beta_{q(\sigma_k^2)}\big)-\mbox{digamma}\big(\alpha_{q(\sigma_k^2)}\big)$\,\,;\,\,$\mu_{q(1/\sigma_k^2)} \longleftarrow \alpha_{q(\sigma_k^2)}\big/\beta_{q(\sigma_k^2)}$
                                \item[] If $K>1$: $[\balpha_{q(\bw)}]_k\longleftarrow\sum_{i=1}^N\mu_{q(a_{ik})} + \alpha_w$ 
                            \end{itemize}
                            \item[] For $k=1,\ldots,K$:  $\mu_{q(\log(w_{k}))}\longleftarrow\mbox{digamma}\big([\balpha_{q(\bw)}]_k\big)-\mbox{digamma}\big(\sum_{k=1}^{K}[\balpha_{q(\bw)}]_k\big)$
                            \item[] For $i=1,\ldots,N$:
						\begin{itemize}
						\setlength\itemsep{2pt}
						\item[] For $k=1,\ldots,K$:
          				\begin{itemize}
						\setlength\itemsep{2pt}
                            \item[] $\tau_{ik}\longleftarrow\mu_{q(\log(w_k))}-\dfrac{1}{2}\mu_{q(\log(\sigma_k^2))}-\dfrac{1}{2}\log(2\pi)$ 
                            \item[]\qquad\qquad$+\mu_{q(1/\sigma_k^2)}\left\{G\left(\bdeta_{{q(\bbeta_k)}};\bx_i\bx_i^T,\mu_{q(\eta_i)}\bx_i,\mu_{q(\eta_i^2)}\right)\right\}$
                            \end{itemize}
                            \item[] For $k=1,\ldots,K$:   $\mu_{q(a_{ik})}\longleftarrow\exp(\tau_{ik})\big/\sum_{k=1}^K\exp(\tau_{ik})$
                            \item[] $\sigma^2_{q(\eta_i)}\longleftarrow\left(\sum_{j:(i,j)\in\mathcal{S}_{\tiny\mbox{obs}}}\mu_{q(\lambda_j^2)}\mu_{q(1/\psi_j^2)} + \sum_{k=1}^K\mu_{q(a_{ik})}\mu_{q(1/\sigma_k^2)} \right)^{-1}$
    			        \item[] $\mu_{q(\eta_i)}\longleftarrow\sigma^2_{q(\eta_i)}\left\{\sum_{k=1}^K\mu_{q(a_{ik})}\mu_{q(1/\sigma_k^2)}\bx_i^T\bmu_{q(\bbeta_k)}+\sum_{j:(i,j)\in\mathcal{S}_{\tiny\mbox{obs}}}\mu_{q(\lambda_j)}(y_{ij}-\mu_{q(\nu_j)})\mu_{q(1/\psi_j^2)}\right\}$
    			        \item[] $\mu_{q(\eta_i^2)}\longleftarrow\sigma^2_{q(\eta_i)}+\mu_{q(\eta_i)}^2$
                            \end{itemize} 
                    \end{itemize}
					\begin{itemize}
						\setlength\itemsep{2pt}

						\item[] For $j = 1,\ldots,M$:
						\begin{itemize}
						\setlength\itemsep{2pt}
    			        \item[] If $j>1$:
    			        \begin{enumerate}
    			        \item[] $\sigma^2_{q(\lambda_j)}\longleftarrow\left(\sum_{i:(i,j)\in\mathcal{S}_{\tiny\mbox{obs}}} \mu_{q(\eta_i^2)}\mu_{q(1/\psi_j^2)}+\dfrac{1}{\sigma^2_\lambda} \right)^{-1}$
    			        \item[]      
                            $\mu_{q(\lambda_j)}\longleftarrow\sigma^2_{q(\lambda_j)}\left\{\sum_{i:(i,j)\in\mathcal{S}_{\tiny\mbox{obs}}}\mu_{q(\eta_i)}\mu_{q(1/\psi_j^2)}\big(y_{ij} -\mu_{q(\nu_j)}\big)+\dfrac{\mu_{\lambda}}{\sigma^2_{\lambda}}\right\}$
    			        \item[]  
                            $\mu_{q(\lambda_j^2)}\longleftarrow\sigma^2_{q(\lambda_j)}+\mu_{q(\lambda_j)}^2$
    			        \end{enumerate}
                        \item[] $\sigma^2_{q(\nu_j)}\longleftarrow\left(\sum_{i:(i,j)\in\mathcal{S}_{\tiny\mbox{obs}}}\mu_{q(1/\psi_j^2)} + \dfrac{1}{\sigma^2_{\nu}} \right)^{-1}$

                        \item[] $\mu_{q(\nu_j)}\longleftarrow\sigma^2_{q(\nu_j)}\left\{\sum_{i:(i,j)\in\mathcal{S}_{\tiny\mbox{obs}}}\mu_{q(1/\psi_j^2)}\big(y_{ij}-\mu_{q(\lambda_j)}\mu_{q(\eta_i)}\big)+\dfrac{\mu_{\nu}}{\sigma^2_{\nu}}\right\}$    
                        \item[] $\mu_{q(\nu_j^2)} \longleftarrow \sigma^2_{q(\nu_j)} + \mu^2_{q(\nu_j)}$
    			    \item[]     
                        $\mu_{q(1/\psi_j^2)} \longleftarrow \alpha_{q(\psi_j^2)}\Big/\Big\{\beta_{\psi^2}+ \dfrac{1}{2}\sum_{i:(i,j)\in\mathcal{S}_{\tiny\mbox{obs}}}\big(y_{ij}^2+\mu_{q(\nu_j^2)}+\mu_{q(\lambda_j^2)}\mu_{q(\eta_i^2)}-2y_{ij}\mu_{q(\nu_j)}$    \item[]\qquad\qquad\qquad\qquad\qquad$-2y_{ij}\mu_{q(\lambda_j)}\mu_{q(\eta_i)}+2\mu_{q(\nu_j)}\mu_{q(\lambda_j)}\mu_{q(\eta_i)}\big)\Big\}$
					\end{itemize}
					\end{itemize}
					\item[] \textbf{Relevant Output:} $\alpha_{q(\psi_j^2)}$ $\beta_{q(\psi_j^2)}=\alpha_{q(\psi_j^2)}/\mu_{q(1/\psi_j^2)}$, $\mu_{q(\lambda_j)}$, $\sigma^2_{q(\lambda_j)}$, $j = 1,\ldots,M$; $\mu_{q(\eta_i)}$, $\sigma^2_{q(\eta_i)}$, $i = 1,\ldots,N$; $\alpha_{q(\sigma_k^2)}$, $\beta_{q(\sigma_k^2)}$, $\bmu_{q(\bbeta_k)}$, $\bSigma_{q(\bbeta_k)}$, $\mu_{q(\log(w_k))}$, $k=1,\ldots,K$.
				\end{itemize}

			\end{small}
		\end{minipage}
	\end{center}
	\caption{\textit{Algorithm for fitting model \eqref{eq:altSEMbis} via MFVB.}}
	\label{alg:MFVBmixtureSEMalt}
\end{algorithm}

\section{Additional results for the real data study}
\label{sec:additionalResultApplication}

\newpage
\begin{figure}
    \centering
    \includegraphics[width=0.8\linewidth]{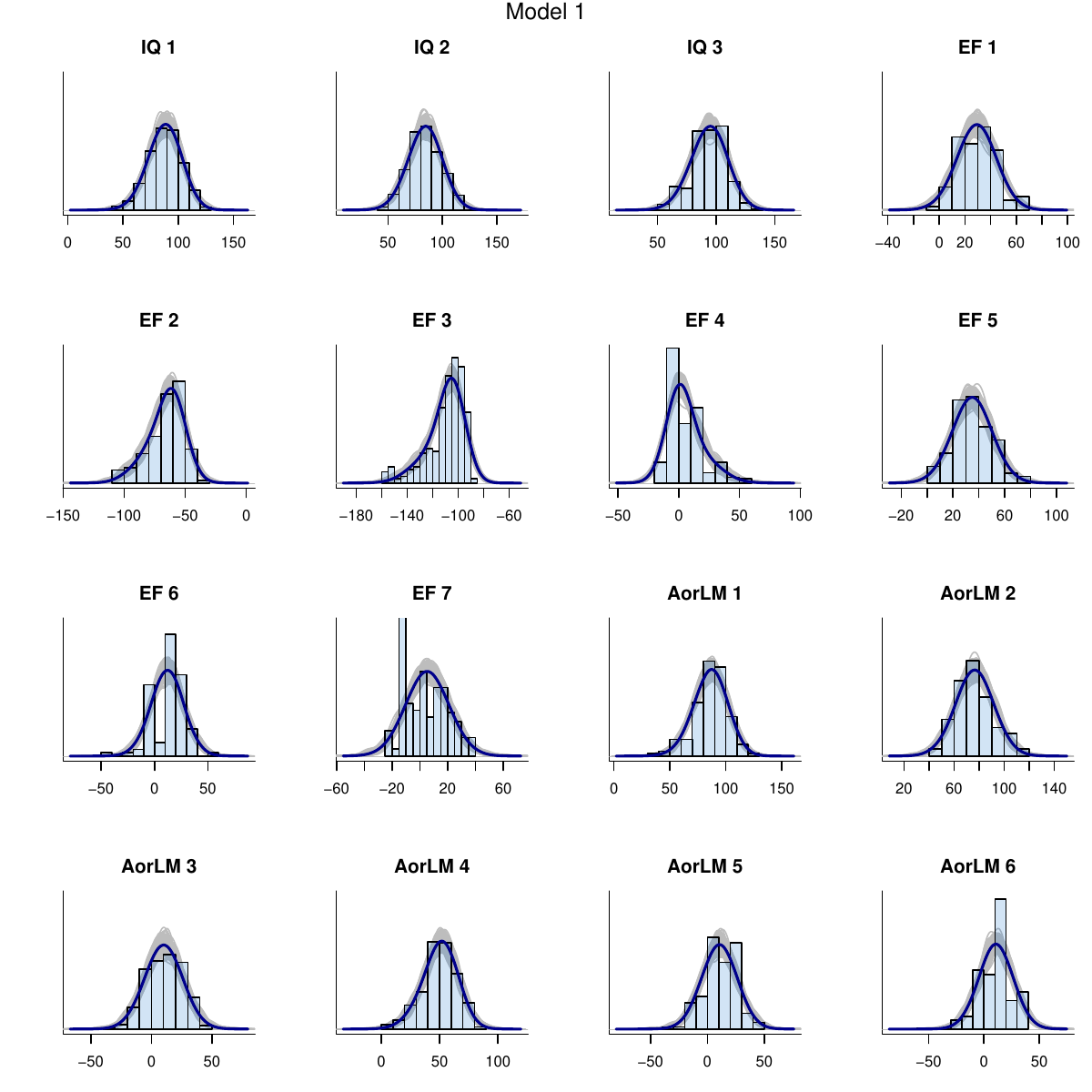}
    \caption{Posterior predictive densities for the outcomes of the Detroit data plotted over histograms of the outcome measurements. Each grey line is the kernel density estimate of one of the $300$ drawings of $\by$ from the posterior predictive distribution from the option 1 fit; their averages are represented as blue lines.}
    \label{fig:ppc_model1}
\end{figure}

\begin{figure}
    \centering
    \includegraphics[width=0.8\linewidth]{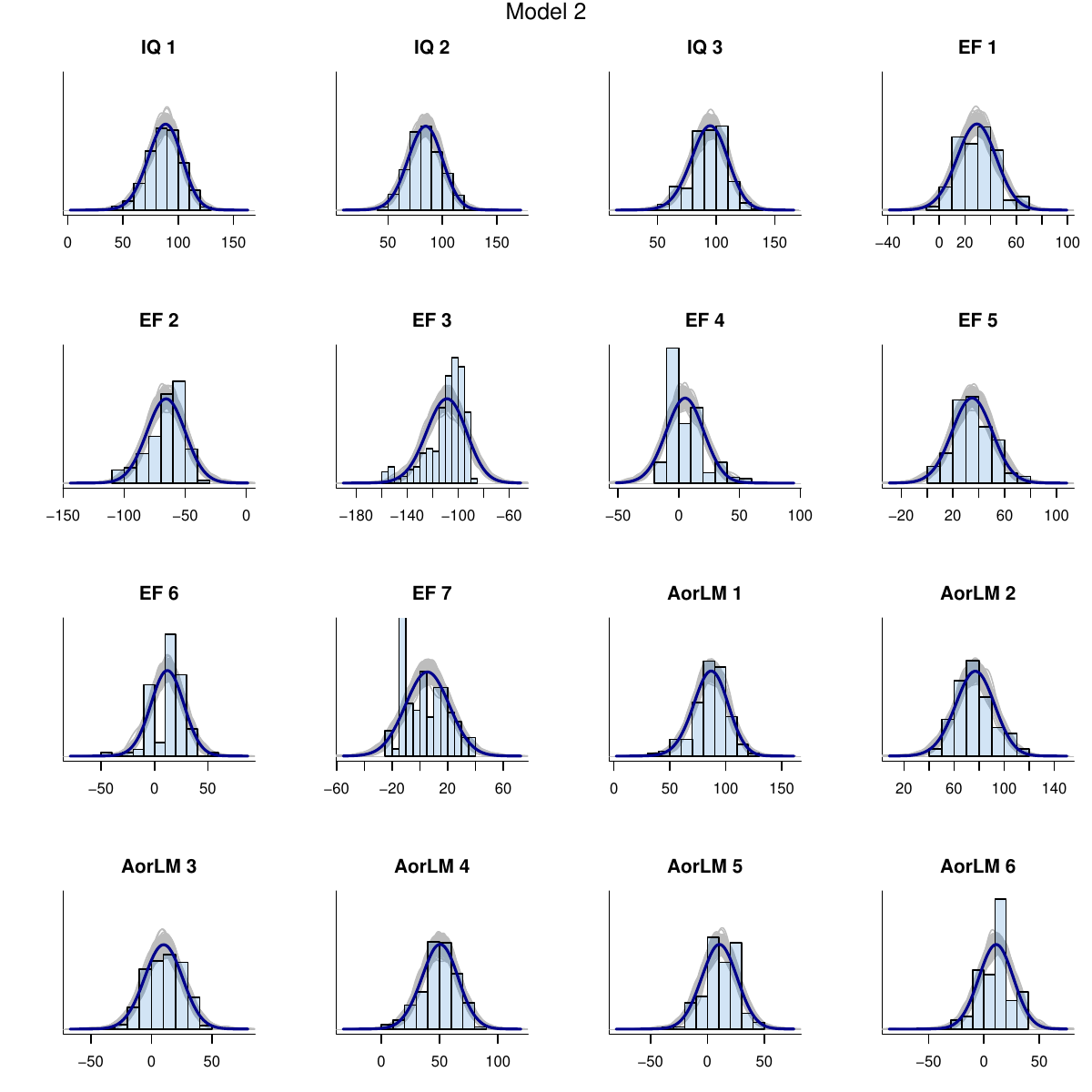}
    \caption{Posterior predictive densities for the outcomes of the Detroit data plotted over histograms of the outcome measurements. Each grey line is the kernel density estimate of one of the $300$ drawings of $\by$ from the posterior predictive distribution from the option 2 fit; their averages are represented as blue lines.}
    \label{fig:ppc_model2}
\end{figure}

\end{document}